\documentclass[journal]{IEEEtran}
\usepackage{times,amsmath,color,amssymb,graphicx,epsfig,cite,psfrag,subfigure,algorithm,balance}
\usepackage{amsfonts,pifont,enumerate,cases}
\usepackage{mathrsfs} 
\usepackage[table]{xcolor} 
\usepackage{verbatim} 
\usepackage{bm}
\usepackage{cuted,stfloats}
\usepackage{algorithmic}
\usepackage{longtable}
\usepackage{blindtext}
\usepackage{multirow}
\usepackage{float}
\usepackage{threeparttable}
\usepackage{makecell}
\usepackage[utf8]{inputenc}
\usepackage{url}
\usepackage{booktabs}
\usepackage{amssymb}
\usepackage{bbding}
\usepackage{pifont}
\usepackage{wasysym}
\usepackage{utfsym}
\usepackage{fontawesome}
\usepackage[algo2e,ruled,linesnumbered,lined,boxed,commentsnumbered]{algorithm2e}

\begin{document}
	
	\title{Cooperative Multi-Static ISAC Networks: A Unified Design Framework for  Active and Passive Sensing}
	\author{Yan Yang,  Zhendong Li,~\IEEEmembership{Member,~IEEE}, Jianwei Zhao, Qingqing Wu,~\IEEEmembership{Senior Member,~IEEE}, \\Zhiqing Wei,~\IEEEmembership{Member,~IEEE}, Wen Chen,~\IEEEmembership{Senior Member,~IEEE}, and Weimin Jia    
			\vspace{-2mm}
			
	\thanks{
		    Y. Yang is with Rocket Force University of Engineering, Xi’an 710025, China, and also with the School of Information and Communication Engineering, Xi’an Jiaotong University, Xi’an 710049, China (e-mail: yangyanyy2022@163.com).
			Z. Li is with the School of Information and Communication Engineering, Xi’an Jiaotong University, Xi’an 710049, China (email: lizhendong@xjtu.edu.cn). 
		    J. Zhao and W. Jia are with Rocket Force University of Engineering, Xi’an 710025, China (e-mail: zhaojianweiep@163.com, jwm602@163.com).
			Q. Wu and W. Chen are with the Department of Electronic Engineering, Shanghai Jiao Tong University, Shanghai 200240, China (e-mail: qingqingwu@sjtu.edu.cn, wenchen@sjtu.edu.cn). 
			Z. Wei is with the Key Laboratory of Universal Wireless Communications, Ministry of Education, Beijing University of Posts and Telecommunications, Beijing 100876, China (e-mail:  weizhiqing@bupt.edu.cn). 
			(Corresponding author: Zhendong Li; Jianwei Zhao)
		}
			
		}
		\maketitle

		\begin{abstract}
			Multi-static cooperative sensing emerges as a promising technology for advancing integrated sensing and communication (ISAC), enhancing sensing accuracy and range. In this paper, we develop a unified design framework for joint active and passive sensing (JAPS). 
			In particular, we consider a JAPS-based cooperative multi-static ISAC system for coexisting downlink (DL) and uplink (UL) communications. An optimization problem is formulated for maximizing the sum rate of both the DL and UL transmissions via jointly optimizing beamforming, receive filters and power allocation, while guaranteeing the sensing requirements and transmission power constraints. However, the formulated problem is a non-convex optimization problem that is challenging to solve directly due to the tight coupling among optimization variables. To tackle this complicated issue, we employ an efficient algorithm architecture leveraging alternating optimization (AO). Specifically, with the given receive filters and transmission power for UL communication, the transmit beamforming subproblem is addressed by successive convex approximation (SCA)-based and penalty-based algorithms. A fractional programming (FP)-based algorithm is developed to tackle the receive filters and transmission power for UL communication optimization subproblem. Extensive numerical results validate the performance improvement of our proposed JAPS scheme and demonstrate the effectiveness of our proposed algorithms.
		\end{abstract}
		
		\begin{IEEEkeywords}
			Integrated sensing and communication, joint active and passive sensing, cooperative multi-static, successive convex approximation, fractional programming.
			
		\end{IEEEkeywords}
		
		\section{Introduction}
		\IEEEPARstart{N}{umerous} emerging applications in the sixth generation (6G) wireless networks, such as smart city, autonomous driving, and intelligent manufacturing \cite{6G magazine,anjiancheng 1}, impose stringent requirements on high-quality wireless data transmission and high-precision sensing capabilities \cite{Proceedings of the IEEE}.  
		As a result, integrated sensing and communication (ISAC) has spurred intensive efforts across both industry and academia communities driven by its potential in improving spectrum, energy and hardware efficiency \cite{Liufan1,anjiancheng 2}.
		It is envisioned to achieve the collaboration and mutual benefit of wireless sensing and communication functions based on shared hardware architectures and spectral resources.
		
		Attracted by the above attractive advantages, there has been extensive literature to explore ISAC systems from different perspectives, e.g., channel estimation \cite{weizhiqing Downlink and Uplink Cooperative, Gao zhen, zhao}, performance analysis \cite{liuyunawei ouyang, anjiancheng, YOLO}, and beamforming design \cite{liuxiang ISAC, liufan CRB, youxiaohu DL UL}. Specifically, in terms of channel estimation, the authors of \cite{weizhiqing Downlink and Uplink Cooperative} designed a unified downlink (DL) and uplink (UL) cooperative ISAC scheme, which used a refined two-dimensional multiple signal classification (MUSIC) algorithm to achieve accurate estimates of the angle of arrival (AoA), range, and velocity.
	In \cite{Gao zhen}, the authors devised a compressive sensing-based channel estimation algorithm for millimeter-wave massive multiple-input multiple-output (MIMO) 
        systems, and provided a processing framework to support target speed measurement and payload data demodulation. Besides, related to performance analysis, the work \cite{liuyunawei ouyang} analyzed the fundamental performance of DL and UL ISAC systems in an information-theoretic viewpoint, and indicated that ISAC can provide more degrees of freedom (DoFs) for both the sensing rate and the communication rate. In \cite{anjiancheng}, the authors discussed the fundamental performance tradeoff between the sensing detection probability and achievable communication throughput in ISAC systems.
		Furthermore, concerning beamforming design, the authors of \cite{liufan CRB} minimized the Cram\'{e}r-rao lower bound (CRLB) for sensing by designing joint transmit beamforming while guaranteeing the communication signal-to-interference-plus-noise ratio (SINR) requirement.
		The authors in \cite{liuxiang ISAC} utilized the weighted sum of independent communication symbols and radar waveforms to form multiple beams, and optimized the transmit beampattern.     
		However, the aforementioned works focused on the mono-static ISAC systems, which did not have enough advantages in sensing range and accuracy as well as communication coverage due to propagation distance and complex obstacles.
		
	   Emerging cooperative multi-static ISAC systems provide a potential solution to break through the limitation of mono-static sensing \cite{PMN ISAC}, which ensures high-quality communication through coordinated signal transmission while enhancing sensing coverage and accuracy via multi-view observations \cite{Multi-cell com, Multi-View Sensing}.
		The compelling demand and potential of cooperative multi-static ISAC networks have motivated a great deal of important studies in ISAC technology.
		Recently, some initial works have been devoted in multi-static ISAC systems. In particular, the authors of \cite{xuchen UAV network} studied a novel beam sharing ISAC scheme based on antenna subarray and corresponding beamforming algorithm to improve the cooperative sensing performance in the unmanned aerial vehicle (UAV) networks.
		In \cite{lihongyu multi-BS}, the authors proposed a practical cooperative sensing method and designed a hybrid beamforming to maximize the system communication capacity.
		In \cite{cell-free ISAC}, a power allocation algorithm was developed to optimize the multi-static sensing SINR in an ISAC-enabled DL cell-free massive MIMO system, while guaranteeing communication requirements. 
		The work \cite{multi-BS DL and UL} optimized the access point duplex mode problem and analyzed the system performance in a cell-free multi-static ISAC network.
		The authors of \cite{yangxiaoyu multi-BS passive} derived the CRLB for estimating target position in the presence of time synchronization errors and proposed a coordinated transmitting beamforming design while satisfying CRLB constraints for sensing.
		Nevertheless, most existing multi-static ISAC works remain confined to either active sensing or passive sensing paradigms, thereby failing to fully leverage the advantages inherent in multi-static cooperation. To be specific, active sensing leverages echoes transmitted from the same base station (BS) for detection, whereas passive sensing extracts target information from reflected signals transmitted from other BSs.
		
       In practical  multi-static ISAC systems, active sensing and passive sensing modalities may coexist. Joint active and passive sensing (JAPS) has emerged as a promising technique,  which takes advantage of the cooperation between coexisting active and passive sensing to achieve more robust environmental adaptability and higher sensing performance \cite{fengyuan gaofeifei}.
		Recently, a limited body of literature has investigated JAPS-based multi-static cooperation technology within ISAC systems.
		For example, the work \cite{Multi—BS} proposed a cooperative ISAC framework for multi-static active and passive sensing, and discussed the key enabling technologies, performance evaluation and research opportunities in detail.
		The authors in\cite{jiangwangjun JAPS} presented a cross-correlation cooperative sensing-based JAPS scheme in the perceptive mobile network having asynchronous transceivers, and developed a low-complexity AoA estimation algorithm adopting coarse and fine precision iterative estimation to realize high-accuracy sensing.
		The authors of \cite{Integrated Active and Passive Sensing} investigated power allocation in a JAPS system with multi-user communications considering the unlimited and limited backhaul capacity cases, respectively.
		Different from \cite{Multi—BS, jiangwangjun JAPS, Integrated Active and Passive Sensing}, the authors in \cite{UL multi-BS} proposed a protocol for UL communications and distributed bi-static sensing.
		However, despite the aforementioned research progress, the above prior works focused on operating in either DL or UL communication. It is extremely crucial to meet both the DL and UL communication demands of the user equipments (UEs) in realistic scenarios.
		In addition, the JAPS works mentioned above made ideal assumptions with perfect self-interference (SI) cancellation at the full-duplex (FD) BS. In practice, SI is an important challenge for the FD systems, which may lead to degrade communication and sensing performances. 
        
		In summary, to fully exploit the performance improvements provided by the cooperative multi-static ISAC networks, the cooperation of active and passive sensing is worth exploring.
		The DL and UL transmissions are tightly coupled in realistic scenarios, but existing ISAC studies usually support only DL or UL communication demands.
		Motivated by the above considerations, this paper investigates an advanced JAPS-based cooperative multi-static ISAC system that simultaneously conducts DL and UL communications and target sensing. 
		Due to the various complex mutual interferences in this scenario, interference management, collaborative
		beamforming, and the tradeoff between sensing and communication are highly complicated and inherently challenging, necessitating dedicated exploration.
		In particular, the main contributions of this paper are detailed as follows:
		\begin{itemize}
			\item{First, we propose a unified design framework for active and passive sensing. 
				The framework introduces a novel JAPS-based cooperative multi-static ISAC system for co-existing DL and UL communications. Specifically, we derive the corresponding system model, and then derive the closed-form expressions for the communication and sensing SINR to evaluate the performance of the multi-UE communication and sensing operation, respectively. Besides, we theoretically prove that the detection probability grows proportionally with sensing SINR when maintaining a constant probability of false alarm.}
			\item{Then, we formulate an optimization problem to maximize the multi-UE sum rate for coupled DL and UL communication while satisfying sensing requirement via jointly optimizing the transmit beamforming,  receive filter for sensing and UL communication, as well as the transmit power of UL UEs. 
			Due to the non-convexity caused by the tightly coupled optimization variables, solving the resulting problem directly is challenging.	
            To address the complicated problem, we propose an alternating optimization (AO) algorithm by leveraging successive convex approximation (SCA) and fractional programming (FP)  techniques to obtain a high-quality solution.}	
			\item{Finally, we provide various simulations to demonstrate the effectiveness of the proposed schemes. Numerical results demonstrate that the performance of the proposed JAPS scheme is superior to both active-only and passive-only sensing benchmarks. Meanwhile, it is also observed that the proposed algorithms can significantly achieve higher sum rates of both DL and UL communications compared with other benchmark algorithms. Besides, we also present the effect of different BS topologies on the performance of our considered system.}
		\end{itemize}
		
		The remainder of this paper is organized as follows. Section II elaborates the system model formulation, derives the performance metrics for communications and sensing, and formulates the sum rate for both DL and UL communication maximization problem. Section III  proposes corresponding alternating optimization algorithms, and analyzes the convergence and computational complexity of the presented algorithms. 
		In Section IV,  numerical results are provided and discussed. 
		Conclusions are drawn in Section V.
		
		\textit{Notations}: Lower-case letters denote scalars, while bold uppercase and lowercase symbols represent matrices and vectors, respectively.  
		The absolute value of a complex-valued scalar $x$ is denoted by $\left|x\right|$.
		For a matrix $\mathbf{X}$, $\mathbf{X}^{H}$, $\mathrm{Tr}(\mathbf{X})$, $\mathrm{rank}(\mathbf{X})$  and $\left\|\mathbf{X}\right\|$ denote its conjugate transpose, trace, rank and matrix norm, respectively.
		$\left\|\cdot\right\|_*$, $\left\|\cdot\right\|_F$ and $\left\|\cdot\right\|_{2}$ represent the nuclear norm, Frobenius norm and spectral norm of the matrix, respectively.
		$\mathbf{X}\succeq0$ indicates that $\mathbf{X}$ is a positive semidefinite matrix. In addition, $\mathbb{C}^{M\times N}$ is the sets of $M\times N$-dimensional complex matrices. $\mathbf{I}_N$ is the identity matrix of dimension $N$.  $j$ denotes the imaginary unit, i.e., $j^{2}=-1$. 
		Denote $\mathcal{O}(\cdot)$ by the big-O computational complexity notation.
		$\mathbb{E}(\cdot)$ denotes the expectation operator.   
		$\mathrm{Re}(\cdot)$ denotes the real part of the argument.  $(\cdot)^{-1}$ and $(\cdot)^{*}$  represents the inverse and conjugate operations. Finally, $\sigma \sim \mathcal{CN} \left(\mu,\mathbf{C}\right)$ means $\sigma$ follows a complex
		Gaussian distribution with mean $\mu$ and covariance matrix $\mathbf{C}$.

		\section{System Model and Problem Formulation}
\begin{figure}[t]
			\centering
			\includegraphics[width=78mm]{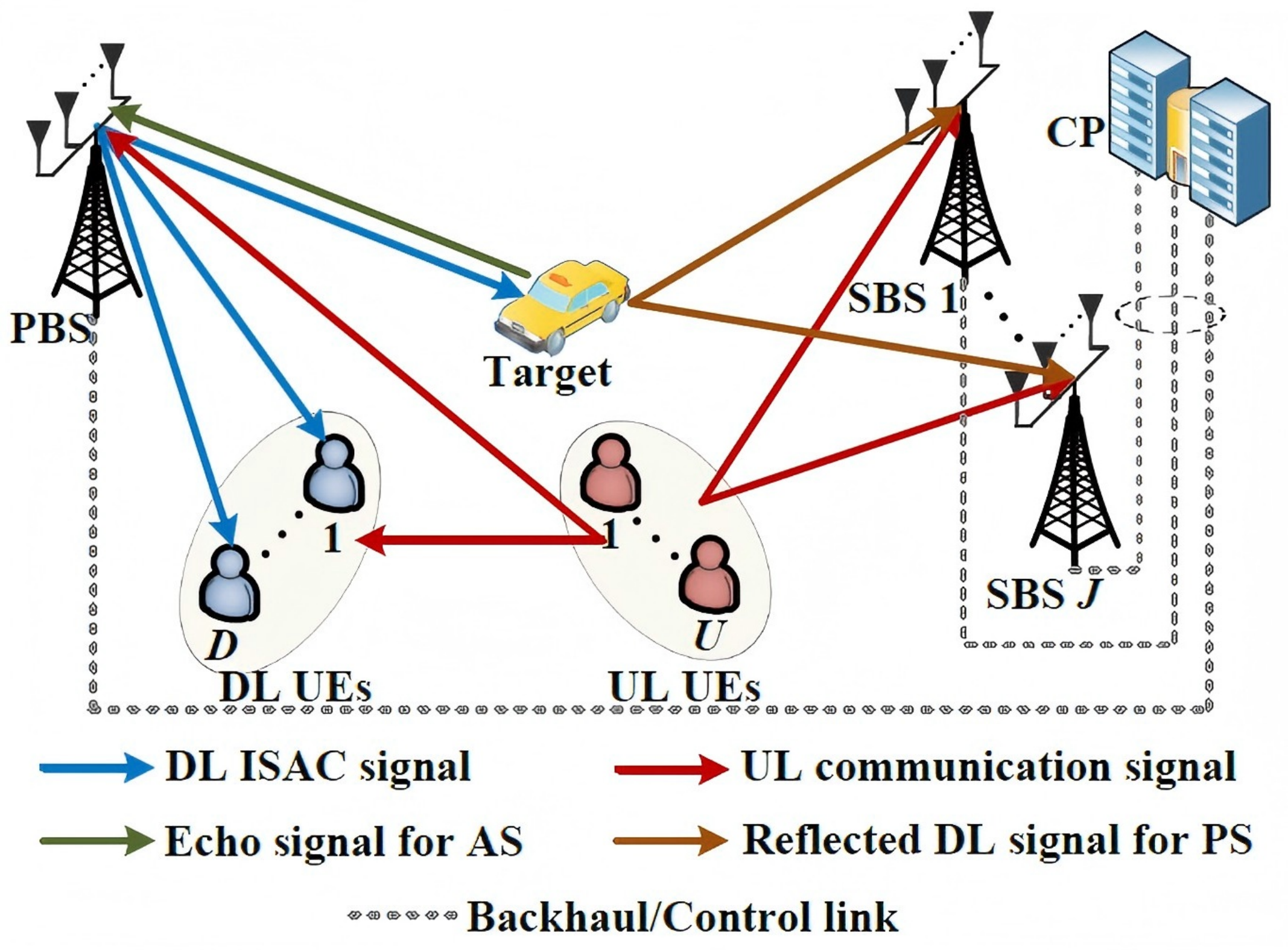}
			\caption{Illustration of cooperative multi-static ISAC networks.}
			\label{fig:multi-static system model}
		\end{figure}

		We consider a JAPS-based cooperative multi-static ISAC system that enables simultaneous target sensing, as well as UL and DL communication, illustrated in Fig.~\ref{fig:multi-static system model}. 
		In this system, a dual-functional FD  primary base station (PBS) is equipped with two separate uniform linear arrays (ULAs) with $M$ transmit antennas and $N_0$ receive antennas, to deliver services to $D$ single-antenna DL UEs, receive the UL communication signals from $U$ single-antenna UL UEs, and simultaneously detect a point target.  Additionally, the system also deploys $J$ secondary base stations (SBSs), each with $N_1$ receive antennas arranged as a ULA, which are designed to simultaneously capture the UL communication signals and the reflected passive sensing signals \footnote{It is worth noting that radar receivers in the traditional radar system do not have prior knowledge of transmitting signals. Different from the radar systems, in the passive sensing between BSs of cooperative multi-static ISAC system, the specific ISAC signal sent by the transmitter is known to the receiver, which is more like  the bi-static sensing \cite{jiangwangjun JAPS}.}. 
	    It is assumed that all BSs are fully synchronized and provide service for multiple UEs via joint transmission, which are connected to a central processor (CP) via backhaul/control links for joint signal processing among the PBS and SBSs \footnote{In practice, for passive sensing, the spatially separated transmitter and receiver are clock-asynchronous, which will lead to time offset and Doppler frequency offset, thereby deteriorating communication and target sensing accuracy, such as causing distance and velocity estimation errors. The works related to clock asynchrony are interesting topics and will be explored as our future work.}.

		\vspace{-2mm}
		\subsection{Transmit Signal Model}
		We denote $\mathbf{s}_\mathrm{c}\in\mathbb{C}^{D\times1}$ as the communication symbols transmitted to the $D$  DL UEs with  $\mathbb{E}\{\mathbf{s}_\mathrm{c}\mathbf{s}_\mathrm{c}^H\}=\mathbf{I}_{D}$, and define $\mathbf{s}_\mathrm{r}\in\mathbb{C}^{M\times1}$ as $M$ individual radar waveforms to extend the DoF of transmit signal satisfying  $\mathbb{E}\{\mathbf{s}_\mathrm{r}\mathbf{s}_\mathrm{r}^{H}\}=\mathbf{I}_{M}$. Assume that $\mathbf{s}_\mathrm{c}$ and $\mathbf{s}_\mathrm{r}$ are independent, i.e., $\mathbb{E}\{\mathbf{s}_\mathrm{c}\mathbf{s}_\mathrm{r}^H\}=\mathbf{0}_{D\times M}$. In addition, we denote the corresponding beamforming matrices for the communication symbols and radar waveforms as $\mathbf{W}_{\mathrm{c}}\in\mathbb{C}^{M\times D}$ and $\mathbf{W}_{\mathrm{r}}\in\mathbb{C}^{M\times M}$, respectively.
		
		For simultaneous support of DL communication and target detection requirements, the dual-functional transmit signal is expressed as 
\begin{equation}\mathbf{x}^{\mathrm{D}}=\mathbf{W}_\mathrm{c}\mathbf{s}_\mathrm{c}^{\mathrm{D}}+\mathbf{W}_\mathrm{r}\mathbf{s}_\mathrm{r}=\mathbf{W}\mathbf{s},\end{equation}
		where $\mathbf{W}\triangleq[\mathbf{W}_\mathrm{c}~\mathbf{W}_\mathrm{r}]\in\mathbb{C}^{M\times(D+M)}$ and  $\mathbf{s}\triangleq[(\mathbf{s}^{\mathrm{D}}_{\mathrm{c}})^{T}~\mathbf{s}_{\mathrm{r}}^{T}]^{T}\in\mathbb{C}^{(D+M)\times1}$ are defined to represent the combined beamforming matrix and symbol vector for brevity, respectively.
		Thus, the covariance matrix of transmit signal can be given by \cite{wuqingqing RIS}
		\begin{equation}
			\mathbf{R}_{\mathbf{w}}\!=\!\mathbb{E}\left[\mathbf{x}^{\mathrm{D}}(\mathbf{x}^{\mathrm{D}})^{{H}}\right]\!\!=\!\!\mathbf{W}\mathbf{W}^{{H}}\!=\!\mathbf{W}_{\mathrm{c}}\mathbf{W}_{\mathrm{c}}^{{H}}\!+\!\mathbf{V}_{\mathrm{r}},
		\end{equation}
		where $\mathbf{V}_{\mathrm{r}} \triangleq \mathbf{W}_{\mathrm{r}}\mathbf{W}_{\mathrm{r}}^{{H}}$ is a general-rank positive semidefinite matrix.
		
		Then, the transmit signal of  UL UE $u$ can be given as  
		\begin{equation}
			x_{u}^{\mathrm{U}}=\sqrt{p_{u}}s_{u}^{\mathrm{U}},\forall u, 
		\end{equation}
		where $0\leq p_{u}\leq P_{u}^\mathrm{max}$ is the transmit power of UL  UE $u$ with $P_{u}^\mathrm{max}$ being the maximum power budget, and $s_{u}^{\mathrm{U}}$ is the UL transmission signal from  UL  UE $u$ to the PBS \cite{youxiaohu DL UL, AISAC}.
		
		\subsection{Sensing Model}
		While transmitting $\mathbf{x}^{\mathrm{D}}$, the PBS simultaneously gathers the UL communication signal and the reflections from the target.
		In the JAPS-based cooperative ISAC framework, the DL communication signals serve dual purposes in communication and sensing, facilitated by the SBSs' complete knowledge of the transmitted symbols.
		Hence, the communication signal is not regarded as interference at the PBS and SBSs.
Without loss of generality, we assume that BS-target links are line-of-sight (LoS) for sensing, given that significant path loss can occur in reflected signals from non-line-of-sight (NLoS) paths.
        
		\subsubsection{Active Sensing Model}
		For active sensing at the PBS, the received signals consist of UL communication signals, desired target reflection, and residual SI, which can be denoted by
		\begin{align}
			\begin{aligned}\label{AS model}
				\mathbf{r}_0=\alpha_0\mathbf{a}_{\mathrm{r},0}(\theta)\mathbf{a}_\mathrm{t}^H(\theta)\mathbf{x}^{\mathrm{D}}+\mathbf{H}_{\mathrm{SI}}\mathbf{x}^{\mathrm{D}}
				+\sum_{u=1}^{U}\mathbf{h}_{0,u}^{\mathrm{U}}x_{u}^{\mathrm{U}}+\mathbf{n}_0,
			\end{aligned}
		\end{align}where $\mathbf{n}_{0}\sim\mathcal{CN}(0,\sigma_{\mathrm{s}}^{2}\mathbf{I}_{N_{0}})$  represents the additive white Gaussian noise (AWGN) vector, $\alpha_0$ denotes the complex sensing coefficient containing the radar cross section (RCS) of the target and the path-loss, and $\theta$  is the detection angle of the target at the PBS. 
		$\mathbf{h}_{0,u}^{\mathrm{U}}$ is the communication channel from UL UE $u$ to the PBS.
		The transmit and receive steering vectors of the antenna array of the PBS can be respectively given as
		\begin{equation}
			\mathbf{a}_{\mathrm{t}}(\cdot)=\frac1{\sqrt{M}}\!\begin{bmatrix}1,e^{j2\pi\Delta\sin(\cdot)},\ldots,e^{j2\pi(M-1)\Delta\sin(\cdot)}\end{bmatrix}^T,
		\end{equation}
		\begin{equation}
			\mathbf{a}_{\mathrm{r},0}(\cdot)=\frac1{\sqrt{N_0}}\!\begin{bmatrix}1,e^{j2\pi\Delta\sin(\cdot)},\ldots,e^{j2\pi(N_0-1)\Delta\sin(\cdot)}\end{bmatrix}^T,
		\end{equation}
		respectively, where $\Delta$ denotes the normalized spacing between adjacent antennas.
		The second term of \eqref{AS model} $\mathbf{H}_{\mathrm{SI}}\mathbf{x}^{\mathrm{D}}$ denotes the residual SI signal. 
		According to \cite{youxiaohu DL UL}, we model the residual SI channel as $\mathbf{H}_{\mathrm{SI}}(l,m)=\sqrt{\beta_{\mathrm{SI}}}e^{-j2\pi r_{l,m}/\lambda}\in\mathbb{C}^{N_0 \times M }$,
		where $\beta_{\mathrm{SI}}$ and $r_{l,m}$ represent the power of residual SI and the space between transmit antenna $l$ and receive antenna $m$, respectively.

		\subsubsection{Passive Sensing Model}
		For passive sensing, the received reflected signal  at the SBS $j$ can be denoted by
		\setcounter{equation}{6}
		\begin{align}
			\begin{aligned}
				\mathbf{r}_j=\alpha_j\mathbf{a}_{\mathrm{r},1}(\varphi_j)\mathbf{a}_{\mathrm{t}}^H(\theta)\mathbf{x}^{\mathrm{D}}+\mathbf{G}_j\mathbf{x}^{\mathrm{D}}+\sum_{u=1}^{U}\mathbf{h}_{j,u}^{\mathrm{U}}x_{u}^{\mathrm{U}}+\mathbf{n}_j,
			\end{aligned}
		\end{align}
		where $\mathbf{n}_{j}\sim\mathcal{CN}(0,\sigma_{\mathrm{s}}^{2}\mathbf{I}_{N_{1}})$ denotes the AWGN vector. $\alpha_j$ is the complex sensing coefficient which follows the model similar to $\alpha_0$.
		$\varphi_j$ is the AoA from the target to SBS $j$. $\mathbf{G}_j\in\mathbb{C}^{N_{1}\times M}$ denotes the direct target-free channel from the PBS to SBS $j$, which is assumed to follow the same channel model as for communication as shown in the following subsection. $\mathbf{h}_{j,u}^{\mathrm{U}}$ is the communication channel from UL UE $u$ to SBS $j$. $\mathbf{a}_{\mathrm{r},1}(\cdot)$ denotes the receive steering vector of each SBS, which is given by
		\begin{equation}
			\mathbf{a}_{\mathrm{r},1}(\cdot)=\frac1{\sqrt{N_1}}\begin{bmatrix}1,e^{j2\pi\Delta\sin(\cdot)},\ldots,e^{j2\pi(N_1-1)\Delta\sin(\cdot)}\end{bmatrix}^T.
		\end{equation}
		It is reasonable to assume that the CP inherently possesses prior knowledge of both the transmit signal $\mathbf{x}^{\mathrm{D}}$ and channel $\mathbf{G}_j$, which can be acquired through the advanced estimation algorithms \cite{weizhiqing Downlink and Uplink Cooperative,Gao zhen,luohongliang Clutter Environment} \footnote{
        In fact, due to limitations in system hardware and the finite beam width, channel estimation may be inaccurate. Although our proposed algorithm is based on the assumption of perfect channel estimation, it can still provide useful performance upper bounds for practical scenarios. To avoid diverting attention from the main focus of this work, the design and analysis of the algorithm under imperfect angles knowledge and channel state information will be addressed in our future work.}. 
		Then,  $\mathbf{r}_j$ can be expressed as
		\begin{align}\label{equ:echo received1}
			\begin{aligned}
				\mathbf{r}_j
				&=\mathbf{A}_j\mathbf{x}^{\mathrm{D}}+\sum_{u=1}^{U}\mathbf{h}_{j,u}^{\mathrm{U}}x_{u}^{\mathrm{U}}+\mathbf{n}_j,
			\end{aligned}
		\end{align}
		where $\mathbf{A}_{j}=\alpha_j\mathbf{a}_{\mathrm{r},1}(\varphi_{j})\mathbf{a}_{\mathrm{t}}^{H}(\theta)$ is preknown at the CP. Similarly, we also define $\mathbf{A}_0=\alpha_0\mathbf{a}_{\mathrm{r},0}(\theta)\mathbf{a}_{\mathrm{t}}^H(\theta)$.
		
		\vspace{-2mm}
		\subsection{Communication Model}
	    The communication channel $\mathbf{h}_{d}^{\mathrm{D}}$ from PBS to the DL UE $d$ and $\mathbf{h}_{\iota,u}^{\mathrm{U}}$ from the UL UE $u$ to PBS/SBSs are assumed to experience both small-scale and large-scale fading and can be respectively formulated as $\mathbf{h}_{d}^{\mathrm{D}}=\sqrt{\beta_d}\bar{\mathbf{h}}_d^{\mathrm{D}},\mathbf{h}_{\iota,u}^{\mathrm{U}}=\sqrt{\beta_{\iota,u}}\bar{\mathbf{h}}_{\iota,u}^{\mathrm{U}}$,
		where $\iota\in\{0,1,2,j,\dots,J\}$, index 0 represents the channel is the communication channel between the UL UE $u$ and the PBS, and index $j$ denotes the channel is the communication channel between the UL UE $u$ and SBS $j$. $\beta_{{d}}=C_0\left(\frac{L_{d}}{L_0}\right)^{-\kappa}$ and $\beta_{\iota,u}=C_0\left(\frac{L_{\iota,u}}{L_0}\right)^{-\kappa}$ represent the large-scale fading coefficients, where $C_0$ is the path-loss at the reference distance $L_0$, $\kappa$ is the path-loss exponent, and $L_{d}$/$L_{\iota,u}$ denotes the corresponding link distance.  $\bar{\mathbf{h}}_d^{\mathrm{D}}$ and  $\bar{\mathbf{h}}_{\iota,u}^{\mathrm{U}}$ are the small-scale fading matrices, which are assumed  to follow the classic Rician fading model as 
		\begin{subequations} 
			\begin{align}
				\setlength{\abovedisplayskip}{3pt}
				\setlength{\belowdisplayskip}{3pt}
				&\bar{\mathbf{h}}_d^{\mathrm{D}}=\sqrt{\frac{\kappa_d}{\kappa_d+1}}\mathbf{h}_{d}^\mathrm{{\mathrm{D}},LoS}+\sqrt{\frac{1}{\kappa_d+1}}\mathbf{h}_{d}^{\mathrm{{\mathrm{D}},NLoS}},\\
				&\bar{\mathbf{h}}_{\iota,u}^{\mathrm{U}}=\sqrt{\frac{\kappa_{\iota,u}}{\kappa_{\iota,u}+1}}\mathbf{h}_{\iota,u}^\mathrm{{\mathrm{U}},LoS}+\sqrt{\frac{1}{\kappa_{\iota,u}+1}}\mathbf{h}_{\iota,u}^{\mathrm{{\mathrm{U}},NLoS}},
			\end{align}
		\end{subequations} 
		where $\kappa_d$, $\kappa_{\iota,u}\geq 0$ are the Rician factors,  capturing the proportion of the energy in the LoS link relative to the energy of the NLoS links.
		In particular, the channel matrix corresponding to the LoS path $\mathbf{h}_{d}^\mathrm{\mathrm{D},LoS}$ and $\mathbf{h}_{\iota,u}^\mathrm{{\mathrm{U}},LoS}$ can be respectively given by 
		\begin{subequations} 
			\begin{align}
				&\mathbf{h}_{d}^\mathrm{{D},LoS}=\mathbf{a}_{\mathrm{t}}({\theta_{d}})\in\mathbb{C}^{M\times1},\\
				&\mathbf{h}_{0,u}^\mathrm{{U},LoS}=\mathbf{a}_{\mathrm{r},0}({\theta_{0,u}})\in\mathbb{C}^{N_0\times1},\\
				&\mathbf{h}_{j,u}^\mathrm{{U},LoS}=\mathbf{a}_{\mathrm{r},1}({\theta_{j,u}})\in\mathbb{C}^{N_1\times1},
			\end{align}
		\end{subequations} 
		where $\theta_d$, $\theta_{0,u}$ and $\theta_{j,u}$ are the directions-of-arrival (DOAs) from PBS to DL UE $d$, from UL UE $u$ to PBS, from UL UE $u$ to SBS $j$. 
		The NLoS Rayleigh fading component $\mathbf{h}_{d}^{\mathrm{{D},NLoS}}$ and $\mathbf{h}_{\iota,u}^{\mathrm{{U},NLoS}}$ follow the distribution with zero mean and unit covariance, i.e., $\mathbf{h}_{d}^{\mathrm{{DL},NLoS}}\sim{\mathcal{CN}}(0,\mathbf{I}_{M})$,  $\mathbf{h}_{0,u}^{\mathrm{{U},NLoS}}\sim{\mathcal{CN}}(0,\mathbf{I}_{N_0})$  and $\mathbf{h}_{j,u}^{\mathrm{{U},NLoS}}\sim{\mathcal{CN}}(0,\mathbf{I}_{N_1})$.
		
		Thus, the received signal at the DL UE $d$ and the combined signal at the SBS and PBS $j$ can be respectively expressed as     
		\begin{align}\label{equ:DL communication}
			\begin{aligned}
				y_{d}^{\mathrm{D}}
				&=(\mathbf{h}_{d}^{\mathrm{D}})^{H}\mathbf{w}_{\mathrm{c},d}s_{\mathrm{c},d}^{\mathrm{D}}+\sum_{d'\neq d}^{D}(\mathbf{h}_{d}^{\mathrm{D}})^{H}\mathbf{w}_{\mathrm{c},d'}s_{\mathrm{c},d'}^{\mathrm{D}}\\
				&+(\mathbf{h}_{d}^{\mathrm{D}})^{H}\mathbf{W}_{\mathrm{r}}\mathbf{s}_{\mathrm{r}}+\sum_{u=1}^{U}h_{d,u}^{\mathrm{du}}x_{u}^{\mathrm{U}}+n_{d}^{\mathrm{D}},
			\end{aligned}
		\end{align}
		\begin{align}\label{equ:UL communication1}
			\begin{aligned}
				y_{0,u}^{\mathrm{U}}
				&=\mathbf{v}_{0,u}^{H}\mathbf{h}_{0,u}^{\mathrm{U}}x_{u}^{\mathrm{U}}+\mathbf{v}_{0,u}^{H}\sum_{u'\neq u}^{U}\mathbf{h}_{0,u'}^{\mathrm{U}}x_{u'}^{\mathrm{U}}\\
				&+\mathbf{v}_{0,u}^{H}\mathbf{A}_0\mathbf{x}^{\mathrm{D}}+\mathbf{v}_{0,u}^{H}\mathbf{H}_{\mathrm{SI}}\mathbf{x}^{\mathrm{D}}+\mathbf{v}_{0,u}^{H}\mathbf{n}_0^{\mathrm{U}},
			\end{aligned}
		\end{align}
		\begin{align}\label{equ:UL communication2}
			\begin{aligned}
				y_{j,u}^{\mathrm{U}}
				&=\mathbf{v}_{j,u}^{H}\mathbf{h}_{j,u}^{\mathrm{U}}x_{u}^{\mathrm{U}}+\mathbf{v}_{j,u}^{H}\sum_{u'\neq u}^{U}\mathbf{h}_{j,u'}^{\mathrm{U}}x_{u'}^{\mathrm{U}}\\
				&+\mathbf{v}_{j,u}^{H}\mathbf{A}_j\mathbf{x}^{\mathrm{D}}+\mathbf{v}_{j,u}^{H}\mathbf{n}_j^{\mathrm{U}},
			\end{aligned}
		\end{align}
		where $\mathbf{v}_{0,u}\in\mathbb{C}^{{N_0}\times1}$ and $\mathbf{v}_{j,u}\in\mathbb{C}^{{N_1}\times1}$ are the receive beamforming vector. $h_{d,u}^\mathrm{du}$ represents the channel from the UL UE $u$ to the DL UE $d$. $\mathbf{s}_\mathrm{c}^{\mathrm{D}}=\begin{bmatrix}s_{\mathrm{c},1}^{\mathrm{D}},s_{\mathrm{c},2}^{\mathrm{D}},\ldots,s_{\mathrm{c},D}^{\mathrm{D}}\end{bmatrix}$.
		$n_{d}^{\mathrm{D}}\sim\mathcal{CN}(0,\sigma_{\mathrm{D}}^{2})$, $\mathbf{n}_0^{\mathrm{U}}\sim\mathcal{CN}(0,\sigma_{\mathrm{U}}^{2}\mathbf{I}_{N_0})$ and $\mathbf{n}_j^{\mathrm{U}}\sim\mathcal{CN}(0,\sigma_{\mathrm{U}}^{2}\mathbf{I}_{N_1})$ are
		the AWGN at DL UE $d$, PBS and  SBS $j$, respectively.
		$\mathbf{w}_{\mathrm{c},j}$ is the $j$-th column of $\mathbf{W}_\mathrm{c}$, i.e., $\mathbf{W}_\mathrm{c}=\begin{bmatrix}\mathbf{w}_{\mathrm{c},1},\mathbf{w}_{\mathrm{c},2},\ldots,\mathbf{w}_{\mathrm{c},D}\end{bmatrix}$.

		\vspace{-3mm}
		\subsection{Active and Passive Sensing Signal Fusion}
		The signals received by the PBS and all SBSs are collected into a vector $\mathbf{r}=\begin{bmatrix}\mathbf{r}_0^T,\mathbf{r}_1^T,\dots,\mathbf{r}_J^T\end{bmatrix}^T$, and the fusion signal is given
		as
		\begin{align}
			\begin{aligned}
				\mathbf{r}\!=\!
				\mathbf{A}\mathbf{W}\mathbf{s}+\mathbf{G}\mathbf{W}\mathbf{s}+\sum_{u=1}^{U}\mathbf{h}_{u}^{\mathrm{U}}x_{u}^{\mathrm{U}}+\mathbf{n},
			\end{aligned}
		\end{align}
		where $\mathbf{G}=[\mathbf{H}_{\mathrm{SI}}^T,\underbrace{\mathbf{0}_{M\times{N_1}},...,\mathbf{0}_{M\times{N_1}}}_{J}]^{T}$ and 
		$\mathbf{h}_{u}^{\mathrm{U}}=[(\mathbf{h}_{0,u}^{\mathrm{U}})^T,(\mathbf{h}_{1,u}^{\mathrm{U}})^T,\ldots,(\mathbf{h}_{J,u}^{\mathrm{U}})^T]^T$.
		$\mathbf{n}=\begin{bmatrix}\mathbf{n}_0^T,\mathbf{n}_1^T,\ldots,\mathbf{n}_J^T\end{bmatrix}^T$ denotes the concatenated noise.
		Based on the above assumptions, $\mathbf{A}$ is known at the CP and can be given by
		\begin{equation}
			\mathbf{A}=[\mathbf{A}_0^T,\mathbf{A}_1^T,...,\mathbf{A}_J^T]^T{\in}\mathbb{C}^{(N_0+JN_1)\times{M}}.
		\end{equation}
		
		\vspace{-3mm}
		\subsection{Performance Metrics}
		
		\subsubsection{Sum Rate for DL and UL Communications}
		The communication SINR should be utilized to characterize the communications performance since the sensing signal interferes negatively with the communications signals in the ISAC system.
		The communication SINR at DL UE $d$ and UL UE $u$ can be respectively computed by \eqref{DL SINR}, as shown on the top of this page, and
		\setcounter{equation}{17}
		\begin{align}
			\begin{aligned}
				&\text{SINR}_u^{\mathrm{U}}
				\!\!=\!\!\frac{p_{u}|\mathbf{v}_{u}^{H}\mathbf{h}_{u}^{\mathrm{U}}|^{2}}{\left|\!\mathbf{v}_{u}^{H}(\sum_{u'\neq u}^{U}\sqrt{p_{u'}}\mathbf{h}_{u'}^{\mathrm{U}})\right|^{2}\!+\!\left\|\!\mathbf{v}_{u}^{H}\mathbf{B}_0\mathbf{W}\right\|_F^{2}\!+\!\sigma_{\mathrm{U}}^{2}\|\mathbf{v}_{u}^{H}\|_{2}^{2}},
			\end{aligned}
		\end{align}
		where $\mathbf{B}_0\triangleq\mathbf{A}+\mathbf{G}$ represents the combined interference channel,
		$\mathbf{v}_{u}=[(\mathbf{v}_{0,u})^T,(\mathbf{v}_{1,u})^T,\ldots,(\mathbf{v}_{J,u})^T]^T\in\mathbb{C}^{(N_0+JN_1)\times{1}}$.
		Then, the achievable sum rate for DL UE $d$ and UL UE $u$ can be respectively expressed as
		\begin{equation}
			R_{d}^{\mathrm{D}}=\log_2\left(1+\text{SINR}_d^{\mathrm{D}}\right),
		\end{equation}
		\begin{equation}
			R_{u}^{\mathrm{U}}=\log_2\left(1+\text{SINR}_u^{\mathrm{U}}\right).
		\end{equation}
		
	\begin{figure*}[ht] 
	\normalsize
	\centering
	\setcounter{equation}{16}
	\begin{align}\label{DL SINR}
			\begin{aligned}
				\text{SINR}_d^{\mathrm{D}}\!=\!=\frac{\left|(\mathbf{h}_{d}^{\mathrm{D}})^{H}\mathbf{w}_{\mathrm{c},d}\right|^2}{\sum_{d'\neq d}^{D}\big|(\mathbf{h}_{d}^{\mathrm{D}})^{H}\mathbf{w}_{\mathrm{c},d'}\big|^2+(\mathbf{h}_{d}^{\mathrm{D}})^{H}\mathbf{V}_{\mathrm{r}}\mathbf{h}_{d}^{\mathrm{D}}+\sum_{u=1}^{U}p_{u}\big|h_{d,u}^{\mathrm{du}}\big|^2+\sigma_{\mathrm{D}}^2}, \forall d,
			\end{aligned}
	\end{align}
	\hrulefill
\end{figure*}

\begin{figure*}[ht] 
	\normalsize
	\centering	
	\setcounter{equation}{25}
	\begin{equation}\label{pd and SINR}
		\begin{aligned}
			P_{\mathrm{D}}\!\propto\!\Omega_{1}/\Omega_{0}
			\!=\!\frac{\mathbb{E}\big\{|\mathbf{u}^H\mathbf{A}\mathbf{W}\mathbf{S}|^{2}\big\}}{\mathbb{E}\big\{|\mathbf{u}^H\mathbf{G}\mathbf{W}\mathbf{S}|^{2}\big\}\!+\!\mathbb{E}\big\{|\mathbf{u}^H\sum_{u=1}^{U}\mathbf{h}_{u}^{\mathrm{U}}\mathbf{x}_{u}^{\mathrm{U}}|^{2}\big\}\!+\!\mathbb{E}\big\{|\mathbf{u}^{H}\mathbf{n}|^{2}\big\}}\!+\!1
			\!=\!\frac{|\mathbf{u}^H\mathbf{A}\mathbf{W}|^{2}}{|\mathbf{u}^H\mathbf{G}\mathbf{W}|^{2}\!+\!\sum_{u=1}^{U}p_{u}|\mathbf{u}^H\mathbf{h}_{u}^{\mathrm{U}}|^{2}\!+\!\sigma_{\mathrm{s}}^{2}\mathbf{u}^{H}\mathbf{u}}\!+\!1,
		\end{aligned}
	\end{equation}
	\hrulefill
\end{figure*}
		
		\subsubsection{SINR for Target Sensing}
		
		Target detection is an important task in sensing. 
		Technically, by applying a receive beamformer $\mathbf{u}\!\triangleq\!\begin{bmatrix}\mathbf{u}_0^T,\mathbf{u}_1^T,\ldots,\mathbf{u}_J^T\end{bmatrix}^T\!\in\!\mathbb{C}^{(N_0+JN_{1})\times 1}$ on the received signal $\mathbf{r}$, we can further enhance the performance of capturing target reflected signals,  where $\mathbf{u}_0\in \mathbb{C}^{N_0\times 1}$ and $\mathbf{u}_j \in \mathbb{C}^{N_{1} \times 1}$. Then,  the received signals at CP for target detecting can be obtained as
		\setcounter{equation}{20}
		\begin{equation}
			\mathrm{r}=\mathbf{u}^H\mathbf{r}=\mathbf{u}^H\mathbf{A}\mathbf{W}\mathbf{s}+\mathbf{u}^H\mathbf{G}\mathbf{W}\mathbf{s}+\mathbf{u}^H\sum_{u=1}^{U}\mathbf{h}_{u}^{\mathrm{U}}{x}_{u}^{\mathrm{U}}+\mathbf{u}^H\mathbf{n}.
		\end{equation}

	   In addition, the target detection procedure can be modeled as a binary hypothesis testing problem (i.e. $\mathcal{H}_1$, target present, or $\mathcal{H}_0$, target absent), which is given by \cite{anjiancheng}
		\begin{equation}
			\begin{cases}\!\mathcal{H}_0\!:\!\mathrm{r}\!=\!\mathbf{u}^H\mathbf{G}\mathbf{W}\mathbf{S}\!+\!\mathbf{u}^H\sum_{u=1}^{U}\mathbf{h}_{u}^{\mathrm{U}}\mathbf{x}_{u}^{\mathrm{U}}\!+\!\mathbf{u}^H\mathbf{n},\\\!\mathcal{H}_1\!:\!\mathrm{r}\!=\!\mathbf{u}^H\mathbf{A}\mathbf{W}\mathbf{S}\!+\!\mathbf{u}^H\mathbf{G}\mathbf{W}\mathbf{S}\!+\!\mathbf{u}^H\sum_{u=1}^{U}\mathbf{h}_{u}^{\mathrm{U}}\mathbf{x}_{u}^{\mathrm{U}}\!+\!\mathbf{u}^H\mathbf{n}.\end{cases}
		\end{equation}
		
         The corresponding conditional probability distributions can be represented as 
		\begin{equation}
			\mathrm{r}\sim\begin{cases}\mathcal{H}_0:\mathcal{CN}(0,\Omega_0),\\\mathcal{H}_1:\mathcal{CN}(0,\Omega_1),\end{cases}
		\end{equation}
		where $\Omega_{0}=\mathbb{E}\big\{|\mathbf{u}^H\mathbf{G}\mathbf{W}\mathbf{S}|^{2}\big\}+\mathbb{E}\big\{|\mathbf{u}^H\sum_{u=1}^{U}\mathbf{h}_{u}^{\mathrm{U}}\mathbf{x}_{u}^{\mathrm{U}}|^{2}\big\}+\mathbb{E}\big\{|\mathbf{u}^{H}\mathbf{n}|^{2}\big\}$ and $\Omega_{1}=\mathbb{E}\big\{|\mathbf{u}^H\mathbf{A}\mathbf{W}\mathbf{S}|^{2}\big\}+\mathbb{E}\big\{|\mathbf{u}^H\mathbf{G}\mathbf{W}\mathbf{S}|^{2}\big\}+\mathbb{E}\big\{|\mathbf{u}^H\sum_{u=1}^{U}\mathbf{h}_{u}^{\mathrm{U}}\mathbf{x}_{u}^{\mathrm{U}}|^{2}\big\}+\mathbb{E}\big\{|\mathbf{u}^{H}\mathbf{n}|^{2}\big\}$.
		
		According to \cite{Estimation Theory}, the Neyman-Pearson detector for target detection can be formulated as $E=|\mathrm{r}|^{2}~{\stackrel{\mathcal{H}_{1}}{\underset{\mathcal{H}_{0}}{\gtrless}}}~\zeta$,
		where $E$ follows chi-squared distribution with two DoFs, and the detection threshold $\zeta$ can be further determined by $\zeta=\frac{|\Omega_0|^2}{2}\mathcal{F}_{\chi_2^2}^{-1}(1-{P}_{\mathrm{FA}})$
		given the desired false alarm probability ${P}_{\mathrm{FA}}$
		In the sequel, the detection probability $P_\text{D}$ can be determined by \cite{Distributed Multi-Node Cooperative }
		\begin{equation}
			\begin{aligned}
				P_\text{D}=\Pr(E>\zeta|\mathcal{H}_1)=1-\mathcal{F}_{\chi_2^2}(2\zeta/\Omega_1),
			\end{aligned}
		\end{equation}
		where $\mathrm{Pr}(\cdot)$ defines the probability function.
		$\mathcal{F}_{\chi_{2}^{2}}(x)=\frac{1}{\Gamma(f/2)}\int_{0}^{x}t^{f/2-1}e^{-t/2}dt$ and 
		$\mathcal{F}_{\chi_{2}^{2}}^{-1}(x)$
		represent the central chi-squared distribution function and its inverse of a chi-square random variable with two DoFs, respectively.
		$\Gamma(\cdot)$ is the Gamma function and $f = 2$ represents the DoF.
		Then,  for a desired $P_{\mathrm{FA}}$, we can obtain detection probability $P_\text{D}$ as \cite{liurang SNR} 
		\begin{equation}
			P_\mathrm{D}=1-\mathcal{F}_{\chi_2^2}\big(\Omega_0/\Omega_1\mathcal{F}_{\chi_2^2}^{-1}(1-P_\mathrm{FA})\big).
		\end{equation}
		
		\begin{figure}[t]
			\centering
			\includegraphics[width=68mm]{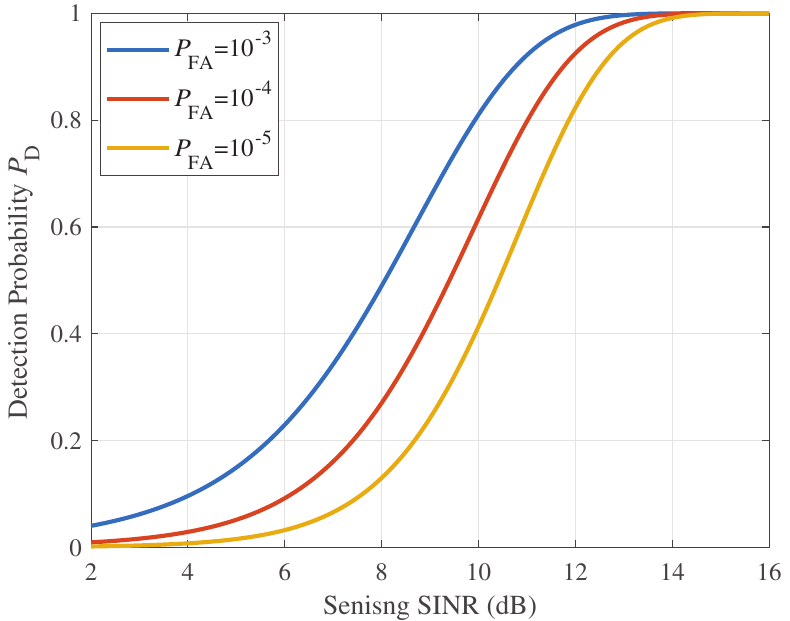}
			\caption{Detection probability versus sensing SINR.}
			\label{fig:Pd_SINR}
		\end{figure}

		Therefore, the relationship of the target detection probability $P_\text{D}$ and the sensing SINR can be derived as \eqref{pd and SINR}.
		The sensing SINR can be written as
		\setcounter{equation}{26}
		\begin{equation}
			\begin{aligned}
				&\text{SINR}_\mathrm{s} =\frac{|\mathbf{u}^H\mathbf{A}\mathbf{W}|^{2}}{|\mathbf{u}^H\mathbf{G}\mathbf{W}|^{2}+\sum_{u=1}^{U}p_{u}|\mathbf{u}^H\mathbf{h}_{u}^{\mathrm{U}}|^{2}+\sigma_{\mathrm{s}}^{2}\mathbf{u}^{H}\mathbf{u}}\\
				&=\frac{\mathbf{u}^H\mathbf{A}\mathbf{W}\mathbf{W}^H\mathbf{A}^H\mathbf{u}}{\mathbf{u}^H(\mathbf{G}\mathbf{W}\mathbf{W}^H\mathbf{G}^H \!+\! \sum_{u=1}^{U}p_{u}\mathbf{h}_{u}^{\mathrm{U}}(\mathbf{h}_{u}^{\mathrm{U}})^H \!+\! \sigma_{\mathrm{s}}^{2}\mathbf{I})\mathbf{u}},
			\end{aligned}
		\end{equation}
		which is positively proportional to the target detection probability $P_\text{D}$ and consequently can be used to evaluate the target detection performance.
		
		To numerically verify the performance of \eqref{pd and SINR}, Fig.~\ref{fig:Pd_SINR} evaluates the effect of different sensing SINRs and false alarm probabilities on the detection probability. 
		It turned out that, given an expected false alarm probability, the target detection performance is positively correlated with the sensing SINR, and the increase of sensing SINR will significantly improve the effective detection probability of the system.
		
		\vspace{-2mm}
		\subsection{Problem Formulation}
		To ensure that the cooperative sensing task accomplishment, beamforming design gives us a way to further improve the target detection performance.
	   Our objective is to maximize the sum rate of all the DL and UL UEs (i.e., $\sum_{d=1}^{D}{R_{d}^{\mathrm{D}}+\sum_{u=1}^{U}R_{u}^{\mathrm{U}}}$) by jointly beamforming and power optimization, while ensuring the sensing SINR requirement as well as transmit power budget at the PBS and UL UEs. We define $\mathcal{A}\triangleq\{\mathbf{u},\mathbf{w}_{\mathrm{c},d},\mathbf{W}_{\mathrm{r}},\mathbf{v}_{u},p_{u},\forall d,u\}$.
		Thus, the optimization problem can be formulated as 
		\begin{subequations}\label{optimization problem P0}
			\begin{align}
				\text{P$_0$:}~\max_{\mathcal{A}}~~&\sum_{d=1}^{D}{R_{d}^{\mathrm{D}}+\sum_{u=1}^{U}R_{u}^{\mathrm{U}}} \label{objective function P0} \\
				\text{s.t.}\quad
				&\text{SINR}_\mathrm{s}\geq\gamma_{\mathrm{s}}, \label{SINR constraint P0}\\
				&\sum_{d=1}^D\|\mathbf{w}_{\mathrm{c},d}\|^2+\mathrm{Tr}(\mathbf{V}_\mathrm{r})\leq P_{\mathrm{max}}^{\mathrm{PBS}},\label{eq:power constraint P0}\\
				&0\leq p_u\leq P_u^{\mathrm{max}},\forall u,\label{UL power}
			\end{align}
		\end{subequations}
		where $\gamma_{\mathrm{s}}$ is the pre-defined sensing SINR threshold, and $P_{\mathrm{max}}^{\mathrm{PBS}}$ denotes the maximum transmission power of the PBS, respectively. 
		The complicated  non-convex problem $\text{P}_0$ is very difficult to solve optimally due to the following reasons: 
		a) the complicated non-concave objective function \eqref{objective function P0} with $\log(\cdot)$ and fractional terms;  
		b) the  coupling among the sensing receive filter $\mathbf{u}$, the transmit beamforming $\mathbf{w}_{\mathrm{c},d}$ and $\mathbf{W}_{\mathrm{r}}$ in the sensing SINR constraint \eqref{SINR constraint P0}  with fractional terms;
		c) the highly coupled variables $\mathbf{w}_{\mathrm{c},d}$, $\mathbf{W}_{\mathrm{r}}$, $\mathbf{v}_{u}$ and $p_{u}$  in the objective function \eqref{objective function P0}. 
		
		\setcounter{equation}{32}
		\begin{figure*}[ht] 
			\normalsize
			\centering	
		\begin{subequations} \label{oumiga DL}   
				\begin{align}
					&\Theta_{d}^{\mathrm{D}}=\operatorname{log}\left(1+\frac{\mathrm{Tr}(\mathbf{h}_{d}^{\mathrm{D}}(\mathbf{h}_{d}^{\mathrm{D}})^{H}\mathbf{W}_{\mathrm{c},d})}{\Psi_{d}^{\mathrm{D}}}\right) \nonumber \\
					&=\operatorname{log}\left(\mathrm{Tr}(\mathbf{h}_{d}^{\mathrm{D}}(\mathbf{h}_{d}^{\mathrm{D}})^{H}\mathbf{W}_{\mathrm{c},d})\! +\! \Psi_{d}^{\mathrm{D}} \right) 
					\!-\!\operatorname{log}\!\bigg(\!\sum_{d'\neq d}^{D}\!\mathrm{Tr}(\mathbf{h}_{d}^{\mathrm{D}}(\mathbf{h}_{d}^{\mathrm{D}})^{H}\mathbf{W}_{\mathrm{c},d'}) \!+\! \mathrm{Tr}(\mathbf{h}_{d}^{\mathrm{D}}(\mathbf{h}_{d}^{\mathrm{D}})^{H}\mathbf{V}_{\mathrm{r}}) \!+\!\sum_{u=1}^{U}p_{u}|{h}_{d,u}^{\mathrm{du}}|^2 \! +\! \sigma_{\mathrm{D}}^{2} \bigg) \label{oumiga DL a} \\
					&\geq\operatorname{log}\!\left(\mathrm{Tr}(\mathbf{h}_{d}^{\mathrm{D}}(\mathbf{h}_{d}^{\mathrm{D}})^{H}\mathbf{W}_{\mathrm{c},d}) \!+\! \Psi_{d}^{\mathrm{D}} \right) 
					\!-\! \bigg(\!a_{d}^{\mathrm{D},{n_1}}\!+\!\sum_{d'\neq d}^{D}\!\mathrm{Tr}(\mathbf{B}_{d}^{\mathrm{D},{n_1}}(\mathbf{W}_{\mathrm{c},d'}\!-\!\mathbf{W}_{\mathrm{c},d'}^{{n_1}})) 
					\!+\!\mathrm{Tr}\big(\mathbf{B}_{d}^{\mathrm{D},{n_1}}(\mathbf{V}_{\mathrm{r}}\!-\!\mathbf{V}_{\mathrm{r}}^{{n_1}})\big) \!\bigg) \triangleq \Theta_{\mathrm{lb},d}^{\mathrm{D}}, \label{oumiga DL b}
				\end{align}
			\end{subequations}
		\end{figure*}
		\begin{figure*}[ht]  
			\vspace{-3mm} 
			\normalsize
			\centering	
		\begin{subequations}\label{oumiga UL}    
				\begin{align}
					&\Theta_{u}^{\mathrm{U}}=\operatorname{log}\bigg(1+\frac{p_{u}|\mathbf{v}_{u}^H\mathbf{h}_{u}^{\mathrm{U}}|^2}{\Psi_{u}^{\mathrm{U}}}\bigg) \nonumber \\
					&=\operatorname{log}\left(p_{u}|\mathbf{v}_{u}^H\mathbf{h}_{u}^{\mathrm{U}}|^2) + \Psi_{u}^{\mathrm{U}}  \right) 
					-\operatorname{log}\bigg(\sum_{u'\neq u}^{U}p_{u'}|\mathbf{v}_{u}^H\mathbf{h}_{u'}^{\mathrm{U}}|^2 + \mathrm{Tr}\big(\mathbf{v}_{u}^H\mathbf{B}_0\mathbf{B}_0^{H}\mathbf{v}_{u}\big(\sum_{d=1}^{D}\mathbf{W}_{\mathrm{c},d}+\mathbf{V}_{\mathrm{r}}\big)\big) + \sigma_{\mathrm{U}}^{2}\|\mathbf{v}_{u}^{H}\|_{2}^{2} \bigg) \label{oumiga UL a} \\
					&\geq\operatorname{log}\left(p_{u}|\mathbf{v}_{u}^H\mathbf{h}_{u}^{\mathrm{U}}|^2) + \Psi_{u}^{\mathrm{U}}  \right) 
					-\bigg(\!a_{u}^{\mathrm{U},{n_1}}+\sum_{d=1}^{D}\mathrm{Tr}(\mathbf{B}_{u}^{\mathrm{U},{n_1}}(\mathbf{W}_{\mathrm{c},d}-\mathbf{W}_{\mathrm{c},d}^{{n_1}})) 
					+\mathrm{Tr}\big(\mathbf{B}_{u}^{\mathrm{U},{n_1}}(\mathbf{V}_{\mathrm{r}}-\mathbf{V}_{\mathrm{r}}^{{n_1}})\big) \!\bigg) \triangleq \Theta_{\mathrm{lb},u}^{\mathrm{U}}, \label{oumiga UL b}
				\end{align}
			\end{subequations}
			\hrulefill
		\end{figure*}
		
		\section{Joint Beamforming Design and Power Optimization in JAPS Framework}
		
		In this section, considering the highly coupling of the variables ${\{\mathbf{u},\mathbf{w}_{\mathrm{c},d},\mathbf{W}_{\mathrm{r}},\mathbf{v}_{u},p_{u},\forall d,u\}}$, we propose an AO algorithm by utilizing penalty-based SCA and FP methods to decouple problem $\text{P}_0$ into three tractable subproblems and find a near-optimal solution.

		\subsection{Receive Filter Design for Sensing}
	    It is noted that the optimization variable $\mathbf{u}$ only exists in the sensing SINR of constraint \eqref{SINR constraint P0} and has no direct influence on the objective function.
		When other optimization variables remain fixed, the original problem $\text{P}_0$ relative to $\mathbf{u}$ is simplified to a standard feasibility check problem, which is given as
		\setcounter{equation}{28}
		\begin{subequations}\label{optimization problem 13}
			\begin{align}
				\operatorname*{find}~&\mathbf{u} \label{objective function 13} \\
				\text{s.t.}
				&\frac{\mathbf{u}^H\mathbf{Q}(\mathbf{W})\mathbf{u}}{\mathbf{u}^H\mathbf{D}(\mathbf{W})\mathbf{u}}\geq\gamma_{\mathrm{s}}, \label{SINR constraint}
			\end{align}
		\end{subequations}
		where  $\mathbf{D}(\mathbf{W})\triangleq \mathbf{G}\mathbf{W}\mathbf{W}^H\mathbf{G}^H \!+\! \sum_{u=1}^{U}p_{u}\mathbf{h}_{u}^{\mathrm{U}}(\mathbf{h}_{u}^{\mathrm{U}})^H \!+\! \sigma_{\mathrm{s}}^{2}\mathbf{I}_{(N_0+JN_1)}$ and $\mathbf{Q}(\mathbf{W})\triangleq\mathbf{A}\mathbf{W}\mathbf{W}^H\mathbf{A}^H$ are the functions with respect to $\mathbf{W}$ that we define for brevity.
		
		In order to ensure sufficient DoFs remain available for subsequent optimization processes and accelerate iterative convergence, we derive the optimal $\mathbf{u}$ by maximizing the sensing SINR. Thus, we have the optimal solution
		\begin{equation} \label{max SINR}
			\mathbf{u}^{\star}=\arg\max_{\mathbf{u}}~
			\frac{\mathbf{u}^H\mathbf{Q}(\mathbf{W})\mathbf{u}}{\mathbf{u}^H\mathbf{D}(\mathbf{W})\mathbf{u}}.
		\end{equation}
		It is obvious that problem \eqref{max SINR} is a generalized Rayleigh maximization, whose optimal solution can be given by the eigenvector associated with the largest eigenvalue of the matrix $(\mathbf{D}(\mathbf{W}))^{-1}\mathbf{Q}(\mathbf{W})$  by applying Rayleigh-Ritz theorem  \cite{rayleigh quotient}.
		
		\vspace{-2mm}
		\subsection{Transmit Beamforming Design for DL Communication and Sensing}\label{Transmit Beamforming Design}
		
	    In this subsection, we focus on the joint transmit beamforming design for given receive filter and UL transmission power based on SCA.
		We start by equivalently transforming the sensing SINR constraint of \eqref{SINR constraint P0} as 
		\begin{equation}	    
			\begin{aligned}
				\mathbf{u}^H\mathbf{A}\mathbf{W}\mathbf{W}^H\mathbf{A}^H\mathbf{u}-\gamma_{\mathrm{s}}\mathbf{u}^H\mathbf{G}\mathbf{W}\mathbf{W}^H\mathbf{G}^H\mathbf{u}-\gamma_{\mathrm{s}}\text{a}_\mathrm{s}\geq 0,
			\end{aligned} \label{SINR constraint transform}
		\end{equation}
		where $\text{a}_\mathrm{s} = \mathbf{u}^H(\sum_{u=1}^{U}p_{u}\mathbf{h}_{u}^{\mathrm{U}}(\mathbf{h}_{u}^{\mathrm{U}})^H)\mathbf{u} \!+\! \sigma_{\mathrm{s}}^{2}\mathbf{u}^H\mathbf{u}$.
		Then, we define $\mathbf{W}_{\mathrm{c},d} = \mathbf{w}_{\mathrm{c},d}\mathbf{w}_{\mathrm{c},d}^{{H}}$ and $\mathcal{W}_{\mathrm{c}}=\{\mathbf{W}_{\mathrm{c},d},\forall d\}$, where $\mathbf{W}_{\mathrm{c},d}\succeq0$ and $\mathrm{rank}(\mathbf{W}_{\mathrm{c},d}) = 1$.	
		Under any given  $\mathbf{u}$, $\{\mathbf{v}_{u},\forall u\}$ and  ${\{p_{u},\forall u\}}$, the original problem $\text{P}_0$ can be converted into 
		\begin{subequations}\label{optimization problem 1}
			\begin{align}
				\text{P$_1$:}~\operatorname*{max}_{\mathcal{W}_{\mathrm{c}},\mathbf{W}_{\mathrm{r}}}~&\sum_{d=1}^{D}\Theta_{d}^{\mathrm{D}}+\sum_{u=1}^{U}\Theta_{u}^{\mathrm{U}} \label{objective function P1} \\
				\text{s.t.}\quad
				&\eqref{SINR constraint transform}, \label{SINR constraint P1}\\
				&\sum_{d=1}^{D}\mathrm{Tr}(\mathbf{W}_{\mathrm{c},d})+\mathrm{Tr}(\mathbf{V}_{\mathrm{r}})\leq P_{\mathrm{max}}^{\mathrm{PBS}},\label{eq:power constraint P1}\\
				&\mathbf{W}_{\mathrm{c},d},\mathbf{V}_{\mathrm{r}}\!\succeq\!0,\mathbf{W}_{\mathrm{c},d},\mathbf{V}_{\mathrm{r}}\!\in\!\mathbb{H}^{M}, \forall d, \label{Wcd,Wr} \\
				&\mathrm{rank}(\mathbf{W}_{\mathrm{c},d})=1,\forall d, \label{rank 1}
			\end{align}
		\end{subequations}
		where 
		$\Theta_{d}^{\mathrm{D}}=\operatorname{log}\left(1+\frac{\mathrm{Tr}(\mathbf{h}_{d}^{\mathrm{D}}(\mathbf{h}_{d}^{\mathrm{D}})^{H}\mathbf{W}_{\mathrm{c},d})}{\Psi_{d}^{\mathrm{D}}}\right)$, $\Theta_{u}^{\mathrm{U}}=\operatorname{log}\bigg(1+\frac{p_{u}|\mathbf{v}_{u}^H\mathbf{h}_{u}^{\mathrm{U}}|^2}{\Psi_{u}^{\mathrm{U}}}\bigg)$, $\Psi_{d}^{\mathrm{D}}=\sum_{d'\neq d}^{D}\mathrm{Tr}(\mathbf{h}_{d}^{\mathrm{D}}(\mathbf{h}_{d}^{\mathrm{D}})^H\mathbf{W}_{\mathrm{c},d'}) + \mathrm{Tr}(\mathbf{h}_{d}^{\mathrm{D}}(\mathbf{h}_{d}^{\mathrm{D}})^H\mathbf{V}_{\mathrm{r}}) + \sum_{u=1}^{U}p_{u}|{h}_{d,u}^{\mathrm{du}}|^2 + \sigma_{\mathrm{D}}^{2}$ and
		${\Psi_{u}^{\mathrm{U}}}=\left|\mathbf{v}_{u}^{H}(\sum_{u'\neq u}^{U}\sqrt{p_{u'}}\mathbf{h}_{u'}^{\mathrm{U}})\right|^{2}\!+\!\left\|\mathbf{v}_{u}^{H}\mathbf{B}_0\mathbf{W}\right\|_F^{2}\!+\!\sigma_{\mathrm{U}}^{2}\left\|\mathbf{v}_{u}^{H}\right\|_{2}^{2}$.
			
		As can be observed, problem $\text{P}_1$ is still non-convex since the objective function \eqref{objective function P1} is non-concave and the rank-one constraint \eqref{rank 1} is non-convex.
		Moreover, the optimization variables  $\mathcal{W}_{\mathrm{c}}$ and $\mathbf{V}_{\mathrm{r}}$ are tightly coupled.  
		Thus, solving problem $\text{P}_1$ is still challenging and needs to be transformed more tractable. 
		
		Next, we employ the SCA method to iteratively transform the objective function \eqref{objective function P1} of problem $\text{P}_1$ into a concave surrogate,  which can be iteratively implemented \cite{xujie SCA}. 
		Define $\{\mathbf{W}_{\mathrm{c},d}^{{n_1}}, \forall d\}$ and $\mathbf{V}_{\mathrm{r}}^{{n_1}}$  as the feasible point obtained in the ${n_1}$-th $({n_1}\geq1)$ iteration, respectively.
        Specifically, we approximate $\Theta_{d}^{\mathrm{D}}$ and $\Theta_{u}^{\mathrm{U}}$ as their lower bound, which follow \eqref{oumiga DL} and \eqref{oumiga UL}, where
		\setcounter{equation}{34}
	\begin{equation}	    
			\begin{aligned}\label{a DL}
				a_{d}^{\mathrm{D},{n_1}}&=\operatorname{log}\bigg(\sum_{d'\neq d}^D\mathrm{Tr}(\mathbf{h}_{d}^{\mathrm{D}}(\mathbf{h}_{d}^{\mathrm{D}})^{H}\mathbf{W}_{\mathrm{c},d'}^{{n_1}})\\& + \mathrm{Tr}(\mathbf{h}_{d}^{\mathrm{D}}(\mathbf{h}_{d}^{\mathrm{D}})^{H}\mathbf{V}_{\mathrm{r}}^{{n_1}})+\sum_{u=1}^{U}p_{u}|{h}_{d,u}^\mathrm{du}|^2 + \sigma_{\mathrm{D}}^{2}\bigg),
			\end{aligned}
		\end{equation}
	\begin{equation}	    
			\begin{aligned}
				a_{u}^{\mathrm{U},{n_1}}&=\operatorname{log}\bigg(\sum_{u'\neq u}^{U}p_{u'}|\mathbf{v}_{u}^H\mathbf{h}_{u'}^{\mathrm{U}}|^2 \\
				&+\!\mathrm{Tr}\bigg(\!\mathbf{v}_{u}^H\mathbf{B}_0\mathbf{B}_0^{H}\mathbf{v}_{u}\bigg(\!\sum_{d=1}^{D}\mathbf{W}_{\mathrm{c},d}^{{n_1}}\!+\!\mathbf{V}_{\mathrm{r}}^{{n_1}}\bigg)\!\bigg) \!\!+\! \sigma_{\mathrm{U}}^{2}\|\mathbf{v}_{u}^{H}\|_{2}^{2}\bigg),
			\end{aligned}
			\label{a UL}
		\end{equation}
		$\mathbf{B}_{d}^{\mathrm{D},{n_1}}$ and $\mathbf{B}_{u}^{\mathrm{U},{n_1}}$ are defined as in \eqref{B DL} and \eqref{B UL}, shown at the top of this page. 
		\begin{figure*}[ht] 
			\vspace{-5mm} 
			\normalsize
			\vspace*{4pt}
			\centering
		\begin{equation}	    
				\begin{aligned}
					\mathbf{B}_{d}^{\mathrm{D},{n_1}}=\frac{\log_2(e)\mathbf{h}_{d}^{\mathrm{D}}(\mathbf{h}_{d}^{\mathrm{D}})^{H}}{\sum_{d'\neq d}^D\mathrm{Tr}(\mathbf{h}_{d}^{\mathrm{D}}(\mathbf{h}_{d}^{\mathrm{D}})^{H}\mathbf{W}_{\mathrm{c},d'}^{n_1}) + \mathrm{Tr}(\mathbf{h}_{d}^{\mathrm{D}}(\mathbf{h}_{d}^{\mathrm{D}})^{H}\mathbf{V}_{\mathrm{r}}^{{n_1}}) +\sum_{u=1}^{U}p_{u}|{h}_{d,u}^{\mathrm{du}}|^2 + \sigma_{\mathrm{D}}^{2}},
				\end{aligned}
				\label{B DL}
			\end{equation}
		\end{figure*}
		\begin{figure*}[ht] 
			\vspace{-5mm} 
			\normalsize
			\vspace*{4pt}
			\centering
			\begin{equation}	    
				\begin{aligned}
					\mathbf{B}_{u}^{\mathrm{U},{n_1}}
					=\frac{\log_2(e)\mathbf{v}_{u}^H\mathbf{B}_0\mathbf{B}_0^{H}\mathbf{v}_{u}}{\sum_{u'\neq u}^{U}p_{u'}|\mathbf{v}_{u}^H\mathbf{h}_{u'}^{\mathrm{U}}|^2 + \mathrm{Tr}\big(\mathbf{v}_{u}^H\mathbf{B}_0\mathbf{B}_0^{H}\mathbf{v}_{u}\big(\sum_{d=1}^{D}\mathbf{W}_{\mathrm{c},d}^{{n_1}}+\mathbf{V}_{\mathrm{r}}^{{n_1}}\big)\big) + \sigma_{\mathrm{U}}^{2}\|\mathbf{v}_{u}^{H}\|_{2}^{2}}.
				\end{aligned}
				\label{B UL}
			\end{equation}
			\hrulefill
		\end{figure*}

		It can be seen that \eqref{oumiga DL a} and \eqref{oumiga UL a}  have concave-minus-concave forms, as well as \eqref{oumiga DL b} and \eqref{oumiga UL b} follow by implementing the first-order Taylor expansion on the second concave term in \eqref{oumiga DL a} and \eqref{oumiga UL a}.
		As a result, we substitute $\Theta_{d}^{\mathrm{D}}$ and $\Theta_{u}^{\mathrm{U}}$ as $\Theta_{\mathrm{lb},d}^{\mathrm{D}}$ and $\Theta_{\mathrm{lb},u}^{\mathrm{U}}$ in  \eqref{objective function P1} of problem $\text{P}_1$, respectively. Therefore, in the $n_1$-th iteration of SCA,
		the problem $\text{P}_1$ is approximated as  
		\begin{subequations}\label{optimization problem P2}
			\begin{align}
				\text{P$_2$:}~\operatorname*{max}_{\mathcal{W}_{\mathrm{c}},\mathbf{W}_{\mathrm{r}}}&\sum_{d=1}^{D}\Theta_{\mathrm{lb},d}^{\mathrm{D}}+\sum_{u=1}^{U}\Theta_{\mathrm{lb},u}^{\mathrm{U}} \label{objective function P3} \\
				\text{s.t.}\quad
				&\eqref{SINR constraint transform},\eqref{eq:power constraint P1},\eqref{Wcd,Wr},\eqref{rank 1}.
			\end{align}
		\end{subequations}
		
		Note that the problem $\text{P}_2$ remains non-convex owing to the inherent non-convexity introduced by the rank-one constraint \eqref{rank 1}.
	To address this obstacle, a widely adopted technique is to utilize the Semidefinite relaxation (SDR). Specifically, the technique firstly ignores the rank-one constraint, and then applies Gaussian randomization or eigenvalue decomposition to derive an approximate solution, in cases where the resulting solution is not of rank-one. However, due to the reconstruction, it may introduce significant performance degradation. Furthermore, given the high dimensionality of the optimization variables, the computational complexity of SDR-based algorithms can become prohibitively high. Hence, we consider applying a double-layer penalty-based iterative algorithm to find a near-optimal rank-one solution \cite{muxidong noma}.
		Toward this idea, the non-convex rank-one constraint \eqref{rank 1} can be equivalently expressed as follows:
		\begin{equation}
			\|\mathbf{W}_{\mathrm{c},d}\|_*-\|\mathbf{W}_{\mathrm{c},d}\|_2=0,\forall d.\label{Wck rank1}
		\end{equation}
	    For any $\mathbf{W}_{\mathrm{c},d}\in\mathbb{H}^{M}$ and $\mathbf{W}_{\mathrm{c},d}\succeq0$, the equality constraint \eqref{Wck rank1} always holds when the matrix $\mathbf{W}_{\mathrm{c},d}$ is rank-one. Otherwise, we must have $\left\|\mathbf{W}_{\mathrm{c},d}\right\|_*-\left\|\mathbf{W}_{\mathrm{c},d}\right\|_2>0$.
		
		For subsequent calculations, we define $\mathfrak{F}(\mathcal{W}_{\mathrm{c}},\mathbf{V}_{\mathrm{r}})=\sum_{d=1}^{D}\Theta_{\mathrm{lb},d}^{\mathrm{D}}+\sum_{u=1}^{U}\Theta_{\mathrm{lb},u}^{\mathrm{U}} $.
		To address the non-convex problem $\text{P}_2$, based on the penalty-based method of \cite{muxidong noma}, we  introduce a penalty factor  $\eta_{1}>0$ and add equality constraint \eqref{Wck rank1} to the objective function \eqref{objective function P3} as a penalty term, yielding  problem $\text{P}_3$ as 
		\begin{subequations}\label{optimization problem P3}
			\begin{align}
				\text{P$_3$:}\operatorname*{max}_{\mathcal{W}_{\mathrm{c}},\mathbf{V}_{\mathrm{r}}}&\mathfrak{F}(\mathcal{W}_{\mathrm{c}},\mathbf{V}_{\mathrm{r}})\!-\!\frac1{\eta_{1}}\!\sum_{d=1}^{D}\!\!\bigg(\!\left\|\mathbf{W}_{\mathrm{c},d}\right\|_*\!-\!\left\|\mathbf{W}_{\mathrm{c},d}\right\|_2\!\!\bigg) \label{objective function P4} \\
				\text{s.t.}\quad
				&\eqref{SINR constraint transform},\eqref{eq:power constraint P1},\eqref{Wcd,Wr}.
			\end{align}
		\end{subequations}
		
		However, for a given $\eta_{1}$, the second term of each penalty term is a concave function in relation to the variable $\mathbf{W}_{\mathrm{c},d}$. Thus, the problem $\text{P}_3$ is still not a convex problem.
		We can solve the optimization problem by employing  SCA  in an alternating manner for a given $\eta_{1}$ until convergence is reached. 
		By utilizing SCA to perform the first-order Taylor expansion at the local point $\mathbf{W}_{\mathrm{c},d}^{n_2}$, its tractable convex upper bound can be derived as
		\begin{equation}
			\begin{aligned}-&\|\mathbf{W}_{\mathrm{c},d}\|_{2}\leq{\mathbf{W}}_{\mathrm{c},d,\mathrm{ub}}^{n_2}\triangleq-\|\mathbf{W}_{\mathrm{c},d}^{n_2}\|_{2}\\&-\mathrm{Tr}\big[\mathbf{v}_{\max}(\mathbf{W}_{\mathrm{c},d}^{n_2})\mathbf{v}_{\max}^{H}(\mathbf{W}_{\mathrm{c},d}^{n_2}) (\mathbf{W}_{\mathrm{c},d}-\mathbf{W}_{\mathrm{c},d}^{n_2}) \big],
			\end{aligned}
		\end{equation}
		where  $\mathbf{W}_{\mathrm{c},d}^{n_2}$ denotes the feasible solution obtained in the ${n_2}$-th iteration, while $\mathbf{v}_{\max}(\mathbf{W}_{\mathrm{c},d}^{n_2})$ represents the eigenvector associated with the largest eigenvalue of $\mathbf{W}_{\mathrm{c},d}^{n_2}$, respectively. Accordingly, problem $\text{P}_3$ can be approximated into the problem $\text{P}_4$, which can be expressed as
		\begin{subequations}\label{optimization problem P4}
			\begin{align}
				\text{P$_4$:}\operatorname*{max}_{\mathcal{W}_{\mathrm{c}},\mathbf{V}_{\mathrm{r}}}&\mathfrak{F}(\mathcal{W}_{\mathrm{c}},\mathbf{V}_{\mathrm{r}})\!-\!\frac1{\eta_{1}}\!\sum_{d=1}^{D}\!\bigg(\!\left\|\mathbf{W}_{\mathrm{c},d}\right\|_*\!+\! {\mathbf{W}}_{\mathrm{c},d,\mathrm{ub}}^{n_2}\!\bigg)  \label{objective function P5} \\
				\text{s.t.}\quad
				&\eqref{SINR constraint transform},\eqref{eq:power constraint P1},\eqref{Wcd,Wr},
			\end{align}
		\end{subequations}
		which is a standard quadratic semidefinite program and can be solved directly by off-the-shelf optimization toolkits \cite{S. Boyd}. 
		
		It bears emphasizing that the selection of appropriate penalty factor $\eta_{1}$ plays a crucial role for the objective function.
		When $\frac{1}{\eta_{1}}\rightarrow+\infty (\eta_{1}\to0)$, we will always have rank-one matrix solutions satisfying the equality constraints \eqref{Wck rank1}.
		To enhance convergence efficiency, the penalty factor $\eta_{1}$ is initialized with a sufficiently large value to secure a favorable starting point, then progressively reduce to a sufficiently small value via $\eta_1=\epsilon_1\eta_1, 0<\epsilon_1<1$, where $\epsilon_1$ denotes a constant scaling coefficient \cite{muxidong noma}. 
		
		For each given $\eta_{1}$, problem $\text{P}_4$ can be addressed in an iterative manner until the fractional diminution of the objective function value is below the convergence threshold $\varepsilon_{1}$ in the inner layer.
		The  proposed algorithm concludes its execution when the equality constraints are fulfilled to the predetermined tolerance threshold, which can be detailed as 
		\begin{equation}
			\max\left\{\left\|\mathbf{W}_{\mathrm{c},d}\right\|_*-\left\|\mathbf{W}_{\mathrm{c},d}\right\|_2, \forall d\right\}\leq\varepsilon_{2},
		\end{equation}
		where $\varepsilon_{2}$ represents the maximum tolerable value. 
        As the penalty factor $\eta_{1}$ decreases, the equality constraints \eqref{Wck rank1} are ultimately satisfied. In each iteration, the subproblem optimizing $\{\mathbf{W}_{\mathrm{c},d}, \forall d\}$ and $\mathbf{V}_{\mathrm{r}}$ attains its optimal solution. Consequently, the objective function of problem \eqref{optimization problem P3} is monotonically non-decreasing over each iteration. Furthermore, due to the sensing SINR requirement and limited transmit power budget at the PBS, the multi-UE sum rate is upper-bounded by a finite value.
		
		\subsection{Receive Filter and  Power Design for UL Transmission}
         In this subsection, we focus on the joint receive filter and UL communication power design for given transmit beamforming by leveraging FP technique.
		In other words, when the other optimization variables $\mathbf{u}$, $\mathcal{W}_{\mathrm{c}}$ and  $\mathbf{V}_{\mathrm{r}}$ are given, the subproblem corresponding to optimize $\{\mathbf{v}_{u},\forall u\}$ and $\{p_{u},\forall u\}$ can be given by
		\begin{subequations}\label{optimization problem P5}
			\begin{align}
				\text{P$_5$:}~\max_{\{\mathbf{v}_{u},p_{u},\forall u\}}~~&\sum_{d=1}^{D}\Theta_{d}^{\mathrm{D}}+\sum_{u=1}^{U}\Theta_{u}^{\mathrm{U}}  \label{objective function P6} \\
				\text{s.t.}\quad
				&\eqref{UL power},\eqref{SINR constraint transform}.\label{SINR constraint P6}
			\end{align}
		\end{subequations}
		
				\begin{figure*}[ht] 
			\vspace{-5mm} 
			\normalsize
			\vspace*{4pt}
			\centering	    
			\begin{align}\label{objective function Lagrangian} 
				&\sum_{d=1}^{D}\!\!\left(\!\operatorname{log}(1\!+\!\delta_{d}^{\mathrm{D}})\!-\!\delta_{d}^{\mathrm{D}}\!+\!\frac{ \big(1 + \delta_{d}^{\mathrm{D}}\big)\mathrm{Tr}(\mathbf{h}_{d}^{\mathrm{D}}(\mathbf{h}_{d}^{\mathrm{D}})^{H}\mathbf{W}_{\mathrm{c},d})}{\sum_{u=1}^{U}p_{u}|{h}_{d,u}^{\mathrm{du}}|^2 + \sum_{d'=1 }^{D}\mathrm{Tr}(\mathbf{h}_{d}^{\mathrm{D}}(\mathbf{h}_{d}^{\mathrm{D}})^{H}\mathbf{W}_{\mathrm{c},d'}) + \mathrm{Tr}(\mathbf{h}_{d}^{\mathrm{D}}(\mathbf{h}_{d}^{\mathrm{D}})^H\mathbf{V}_{\mathrm{r}}) + \sigma_{\mathrm{D}}^{2}}\!\right) \nonumber \\
				&	\!+\!\sum_{u=1}^{U}\!\!\left(\!\operatorname{log}(1\!+\!\delta_{u}^{\mathrm{U}})\!-\!\delta_{u}^{\mathrm{U}}\!+\!\frac{\big(1 + \delta_{u}^{\mathrm{U}}\big)p_{u}|\mathbf{v}_{u}^{H}\mathbf{h}_{u}^{\mathrm{U}}|^{2}}{\sum_{u'=1}^{U}p_{u'}|\mathbf{v}_{u}^{H}\mathbf{h}_{u'}^{\mathrm{U}}|^{2} + \left\|\mathbf{v}_u^H\mathbf{B}_0\mathbf{W}\right\|_F^2\!+\!\sigma_{\mathrm{U}}^2\|\mathbf{v}_{u}^{H}\|_{2}^{2}}\!\right), 
			\end{align}
			\hrulefill
		\end{figure*}
		
		\subsubsection{FP-based Transformation}
      By noting the fact that problem $\text{P}_5$ is a sum-of-functions-of-ratio problem,  
      the Lagrangian dual transform is employed to equivalently convert the objective function \eqref{objective function P6} into \eqref{objective function Lagrangian}, displayed at the top of this page \cite{Fractional programming 1}, where 
		$\boldsymbol{\delta}=\{\delta_{d}^{\mathrm{D}}\geq0,\delta_{u}^{\mathrm{U}}\geq0,\forall d,u\}$ denotes the introduced auxiliary variable.

		Then, to further convert the objective function \eqref{objective function Lagrangian} into a more solvable structure, the quadratic transform is applied, and we can obtain \eqref{objective function quadratic transform}, displayed at the top of next page \cite{Fractional programming 1},
		\begin{figure*}[ht] 
			\vspace{-5mm} 
			\normalsize
			\vspace*{4pt}
			\centering	    
			\begin{align}\label{objective function quadratic transform} 
				&\sum_{d=1}^{D}\!\!\bigg(\!\!\log(1\!+\!\delta_{d}^{\mathrm{D}})\!-\!\delta_{d}^{\mathrm{D}}\!+\!2\sqrt{1\! +\! \delta_{d}^{\mathrm{D}}}\mathrm{Re}\{(\eta_{d}^{\mathrm{D}})^{*}\mathbf{h}_{d}^{\mathrm{D}}\mathbf{w}_{\mathrm{c},d}\}\!-\!|\eta_{d}^{\mathrm{D}}|^{2}\bigg(\!\sum_{u=1}^{U}p_{u}|{h}_{d,u}^{\mathrm{du}}|^2 \!+ \!\sum_{d'=1}^{D}\!\mathrm{Tr}(\mathbf{h}_{d}^{\mathrm{D}}(\mathbf{h}_{d}^{\mathrm{D}})^{H}\mathbf{W}_{\mathrm{c},d'}) \nonumber \\
				& + \mathrm{Tr}(\mathbf{h}_{d}^{\mathrm{D}}(\mathbf{h}_{d}^{\mathrm{D}})^H\mathbf{V}_{\mathrm{r}}) \!+\! \sigma_{\mathrm{D}}^{2}\bigg)\bigg) +\sum_{u=1}^{U}\bigg(\log(1+\delta_{u}^{\mathrm{U}})-\delta_{u}^{\mathrm{U}}
				+2\sqrt{(1 + \delta_{u}^{\mathrm{U}})p_{u}}\mathrm{Re}\{(\eta_{u}^\mathrm{U})^{*}\mathbf{v}_{u}^H\mathbf{h}_{u}^{\mathrm{U}}\} \nonumber  \\
				&-|\eta_{u}^\mathrm{U}|^2\bigg(\sum_{u'=1}^{U}p_{u'}|\mathbf{v}_{u}^{H}\mathbf{h}_{u'}^{\mathrm{U}}|^{2} + \left\|\mathbf{v}_u^H\mathbf{B}_0\mathbf{W}\right\|_F^2+\sigma_{\mathrm{U}}^2\|\mathbf{v}_{u}^{H}\|_{2}^{2}\bigg)\bigg),
			\end{align}
			\hrulefill
		\end{figure*}
		where $\boldsymbol{\eta} = \{\eta_{d}^{\mathrm{D}} \geq 0,\eta_{u}^{\mathrm{U}} \geq 0,\forall d,u\}$ is also an introduced auxiliary variable. 
		
		\subsubsection{Update $\boldsymbol{\delta}$ and $\boldsymbol{\eta}$}
		When $\{\mathbf{v}_{u},\forall u\}$ and $\{p_{u},\forall u\}$ are given, the optimal $\boldsymbol{\delta}^{\star}$ can be obtained in the following closed form by setting the derivative of the objective function \eqref{objective function quadratic transform} in relation to $\boldsymbol{\delta}$ to zero, which can be calculated as
		\begin{subequations}\label{delta star}
			\begin{align}
				&\begin{aligned}	
					&\delta_{d}^{\mathrm{D},\star}\!=\\
					&\!\frac{|\mathbf{h}_{d}^{\mathrm{D}}\mathbf{w}_{\mathrm{c},d}|^2}{\sum_{d'\neq d}^{D}\!|\mathbf{h}_{d}^{\mathrm{D}}\mathbf{w}_{\mathrm{c},d'}|^2\! + \! \mathbf{h}_{d}^{\mathrm{D}}\mathbf{V}_{\mathrm{r}}(\mathbf{h}_{d}^{\mathrm{D}})^H \!+ \!\sum_{u=1}^{U}\!p_{u}|{h}_{d,u}^{\mathrm{du}}|^2 \!+\! \sigma_{\mathrm{D}}^{2}},\forall d,
				\end{aligned} \\
				&\delta_{u}^{\mathrm{U},\star}\!\!=\!\!\frac{p_{u}\left|\mathbf{v}_{u}^H\mathbf{h}_{u}^{\mathrm{U}}\right|^2}{\left|\mathbf{v}_{u}^{H}(\sum_{u'\ne u}^{U}\!\sqrt{p_{u'}}\mathbf{h}_{u'}^{\mathrm{U}})\right|^{2}\!\!+\!\!\left\|\mathbf{v}_{u}^{H}\mathbf{B}_0\mathbf{W}\right\|_F^{2}\!+\!\sigma_{\mathrm{U}}^{2}\|\mathbf{v}_{u}^{H}\|_{2}^{2}},\forall u.	\end{align}
		\end{subequations}
		
		When $\{\mathbf{v}_{u},\forall u\}$, $\{p_{u},\forall u\}$ and $\boldsymbol{\delta}$ are held fixed, by setting the derivative of the objective function \eqref{objective function quadratic transform} in relation to $\boldsymbol{\eta}$ to zero, we can obtain the optimal $\boldsymbol{\eta}^{\star}$  as
		\begin{subequations}\label{eq: eta star}
			\begin{align}
				&\begin{aligned}&\eta_{d}^{\mathrm{D},\star}=\\
					&	\frac{\sqrt{1 + \delta_{d}^{\mathrm{D}}}\mathbf{h}_{d}^{\mathrm{D}}\mathbf{w}_{\mathrm{c},d}}{\sum_{d'=1}^{D}\!|\mathbf{h}_{d}^{\mathrm{D}}\mathbf{w}_{\mathrm{c},d'}|^2\!\! + \!\! \mathbf{h}_{d}^{\mathrm{D}}\mathbf{V}_{\mathrm{r}}(\mathbf{h}_{d}^{\mathrm{D}})^H \!\!+ \!\sum_{u=1}^{U}\!p_{u}|{h}_{d,u}^{\mathrm{du}}|^2\! \!+\!\! \sigma_{\mathrm{D}}^{2}},\!\forall d,
				\end{aligned} \\
				&\eta_{u}^{\mathrm{U},\star}=\frac{\sqrt{(1 + \delta_{u}^{\mathrm{U}})p_{u}}\mathbf{v}_{u}^H\mathbf{h}_{u}^{\mathrm{U}}}{\sum_{u'=1}^{U}p_{u'}|\mathbf{v}_{u}^{H}\mathbf{h}_{u'}^{\mathrm{U}}|^{2}\!+\!\left\|\mathbf{v}_{u}^{H}\mathbf{B}_0\mathbf{W}\right\|_F^{2}\!+\!\sigma_{\mathrm{U}}^{2}\|\mathbf{v}_{u}^{H}\|_{2}^{2}},\forall u.
			\end{align}
		\end{subequations}
		
		\subsubsection{Update $\{\mathbf{v}_{u},\forall u\}$}
		
		Given the other variables $\{p_{u},\forall u\}$, $\boldsymbol{\delta}$ and $\boldsymbol{\eta}$, the optimization for $\{\mathbf{v}_{u},\forall u\}$  is formulated as
		\begin{equation}\label{optimization problem P6}
			\begin{aligned}
				\text{P$_6$:}\min_{\{\mathbf{v}_{u},\forall u\}}~&\sum_{u=1}^{U}\{\mathbf{v}_{u}^{H}\mathbf{\Lambda}_{1,u}\mathbf{v}_{u}-2\mathrm{Re}\{\mathbf{v}_{u}^{H}\mathbf{\lambda}_{1,u}\}\}+\rho_{1}  
			\end{aligned}
		\end{equation}
		where we define the coefficients $\mathbf{\Lambda}_{1,u}$ and $\mathbf{\lambda}_{1,u}$ as follows
		\begin{subequations}
			\begin{align}\label{D1,d1}
				&\mathbf{\Lambda}_{1,u}\!=\!|\eta_{u}^\mathrm{U}|^2\bigg(\!\sum_{u'=1}^{U}p_{u'}\mathbf{h}_{u'}^{\mathrm{U}}(\mathbf{h}_{u'}^{\mathrm{U}})^{H}\! + \!\mathbf{B}_0\mathbf{W}\mathbf{W}^H\mathbf{B}_0^H + \sigma_{\mathrm{U}}^2\mathbf{I}\!\bigg),\\
				&\mathbf{\lambda}_{1,u}=\sqrt{(1 + \delta_{u}^{\mathrm{U}})p_{u}}(\eta_{u}^\mathrm{U})^{*}\mathbf{h}_{u}^{\mathrm{U}},	\end{align}
		\end{subequations}
		and $\rho_{1}$ is a constant term that does not affect problem-solving and is thus omitted due to space limitation.
		
		Note that the problem P$_6$ consists of ${U}$ decoupled sub-problems, so we only need to focus on any one of them, i.e.,
		\begin{equation}\label{optimization problem P7}
			\begin{aligned}
				\text{P$_7$:}~\min_{\mathbf{v}_{u}}~~\mathbf{v}_{u}^{H}\mathbf{\Lambda}_{1,u}\mathbf{v}_{u}-2\mathrm{Re}\{\mathbf{v}_{u}^{H}\mathbf{\lambda}_{1,u}\} 
			\end{aligned}
		\end{equation}
		
		Then, the optimal $\mathbf{v}_{u}^{\star}$ of problem P$_7$ can be obtained in the closed expression by setting the derivative of the objective function \eqref{optimization problem P7} in relation to $\mathbf{v}_{u}$ to zero, which can be given by
		\begin{equation}\label{optimization vu}
			\mathbf{v}_{u}^{\star}=\mathbf{\Lambda}_{1,u}^{-1}\mathbf{\lambda}_{1,u}, \forall u.
		\end{equation}
		
		\subsubsection{Update $\{p_{u},\forall u\}$}
		
		Given the other variables $\{\mathbf{v}_{u},\forall u\}$, $\boldsymbol{\delta}$ and $\boldsymbol{\eta}$, the optimization for $\{p_{u},\forall u\}$  is formulated as
		\begin{subequations}\label{optimization problem P8}
			\begin{align}
				\text{P$_8$:}~\min_{\{p_{u},\forall u\}}~~&\sum_{u=1}^{U}(\mu_{1,u}p_u-\mu_{2,u}\sqrt{p_u})+\rho_{2}  \label{objective function P9} \\
				\text{s.t.}\quad
				&\sum_{u=1}^{U}\mu_{3,u}p_{u}+\rho_{3}\leq0, \label{SINR constraint P9}\\
				&\eqref{UL power},
			\end{align}
		\end{subequations}
		where we define the parameters in problem $\text{P}_8$ as follows:  
		\begin{subequations}
			\begin{align}\label{a1,c2}
				&{\mu}_{1,u}\!=\!\sum_{d=1}^{D}|\eta_{d}^{\mathrm{D}}|^{2}|{h}_{d,u}^{\mathrm{du}}|^2 +|\eta_{u}^\mathrm{U}|^2\sum_{u'=1}^{U}|\mathbf{v}_{u'}^{H}\mathbf{h}_{u}^{\mathrm{U}}|^{2},\\
				&{\mu}_{2,u}=2\sqrt{1 + \delta_{u}^{\mathrm{U}}}\mathrm{Re}\{(\eta_{u}^\mathrm{U})^{*}\mathbf{v}_{u}^H\mathbf{h}_{u}^{\mathrm{U}}\},\\
				&{\mu}_{3,u} = |\mathbf{u}^H\mathbf{h}_{u}^{\mathrm{U}}|^2.		
			\end{align}
		\end{subequations}
		Both $\rho_{2}$ and $\rho_{3}$ are constant terms, where $\rho_{2}$ can be omitted during the optimization, and $\rho_{3}$ can be easily
		 obtained on the basis of \eqref{SINR constraint transform}, which can be given by $\rho_3=\sigma_\mathrm{s}^2\mathbf{u}^H\mathbf{u}+\mathbf{u}^H\mathbf{G}\mathbf{W}\mathbf{W}^H\mathbf{G}^H\mathbf{u}-\frac{1}{\gamma_\mathrm{s}}\mathbf{u}^H\mathbf{A}\mathbf{W}\mathbf{W}^H\mathbf{A}^H\mathbf{u}$.

		Obviously, the problem  $\text{P}_8$  is a simple convex problem that can be directly solved by applying the CVX toolbox \cite{S. Boyd}.
		In the third subproblem $\text{P}_5$, $\{\mathbf{v}_{u},\forall u\}$, $\{p_{u},\forall u\}$, $\boldsymbol{\delta}$ and $\boldsymbol{\eta}$ are alternately optimized in our proposed AO algorithm framework.

		\vspace{-2mm}
		\subsection{Convergence Analysis and Computational Complexity}
		In this section, our proposed optimization algorithm for the  JAPS-based cooperative multi-static ISAC  system is summarized in \textbf{Algorithm \ref{overall algorithm}}.
		\begin{algorithm}
			\renewcommand{\algorithmicrequire}{\textbf{Input:}}
			\renewcommand{\algorithmicensure}{\textbf{Output:}}
			\caption{Joint beamforming design and power optimization algorithm in JAPS Framework.}
			\label{alg1}
			\begin{algorithmic}[1]
				\REQUIRE   iteration number $n=1$ and convergence threshold $\xi$.
				\STATE Initialize feasible points $\mathcal{W}_{\mathrm{c}}^{0}$,  $\mathbf{V}_{\mathrm{r}}^{0}$, $\{\mathbf{v}_{u}^{0},\forall u\}$, $\{p_{u}^{0},\forall u\}$ and the penalty factor $\eta_1$;
				\REPEAT
				\STATE  Update the optimization variables $\mathbf{u}^{n}$ via \eqref{max SINR};
				\STATE Given $\{\mathbf{v}_{u}^{n-1},\forall u\}$ and $\{p_{u}^{n-1},\forall u\}$, update $\mathcal{W}_{\mathrm{c}}^{n}$ and $\mathbf{W}_{\mathrm{r}}^{n}$ via 
				solving problem $\text{P}_4$ by applying SCA-based and  penalty-based  methods;
				\STATE Given $\mathcal{W}_{\mathrm{c}}^{n-1}$ and $\mathbf{V}_{\mathrm{r}}^{n-1}$, update $\{\mathbf{v}_{u}^{n},\forall u\}$ and $\{p_{u}^{n},\forall u\}$ via 
				\eqref{optimization vu} and solving problem $\text{P}_8$ by applying FP method;
				\STATE $n = n + 1$;
				\UNTIL The fractional diminution of the objective function value falls below a predetermined threshold $\xi$. 
				\ENSURE $\mathbf{u}^{\star}$, $\mathcal{W}_{\mathrm{c}}^{\star}$, $\mathbf{V}_{\mathrm{r}}^{\star}$, $\{\mathbf{v}_{u}^{\star},\forall u\}$ and $\{p_{u}^{\star},\forall u\}$.
			\end{algorithmic}\label{overall algorithm}  
		\end{algorithm}
        
		\subsubsection{Convergence Analysis}
		The variables $\mathbf{u}$, $\mathcal{W}_{\mathrm{c}}$, $\mathbf{V}_{\mathrm{r}}$, $\{\mathbf{v}_{u},\forall u\}$ and $\{p_{u},\forall u\}$ are updated in an alternating manner until the sum rate achieves convergence. We define the objective function as $f\big(\mathbf{u}^{n},\mathcal{W}_{\mathrm{c}}^{n},\mathbf{V}_{\mathrm{r}}^{n},\{\mathbf{v}_{u}^{n},\forall u\}, \{p_{u}^{n},\forall u\}\big)$, where $\mathbf{u}^{n}$, $\mathcal{W}_{\mathrm{c}}^{n}$, $\mathbf{V}_{\mathrm{r}}^{n}$, $\{\mathbf{v}_{u}^{n},\forall u\}$ and  $\{p_{u}^{n},\forall u\}$ denote the optimal solutions of the formulated problem in the $n$-th iteration.    
	Based on the above derivations, we can obtain
		\begin{align}
			f&\bigg(\!\!\mathbf{u}^{n},\mathcal{W}_{\mathrm{c}}^{n},\mathbf{V}_{\mathrm{r}}^{n},\{\mathbf{v}_{u}^{n},\forall u\}, \{p_{u}^{n},\forall u\}\!\!\bigg)\nonumber \\
			&\overset{a}{\leq} f\bigg(\!\!\mathbf{u}^{n+1},\mathcal{W}_{\mathrm{c}}^{n},\mathbf{V}_{\mathrm{r}}^{n},\{\mathbf{v}_{u}^{n},\forall u\}, \{p_{u}^{n},\forall u\}\!\!\bigg)\nonumber \\
			&\overset{b}{\leq}  f\bigg(\!\!\mathbf{u}^{n+1},\mathcal{W}_{\mathrm{c}}^{n+1},\mathbf{V}_{\mathrm{r}}^{n+1},\{\mathbf{v}_{u}^{n},\forall u\}, \{p_{u}^{n},\forall u\}\!\!\bigg)\nonumber \\
			&\overset{c}{\leq} f\bigg(\!\!\mathbf{u}^{n+1},\mathcal{W}_{\mathrm{c}}^{n+1},\mathbf{V}_{\mathrm{r}}^{n+1}, \{\mathbf{v}_{u}^{n+1},\forall u\}, \{p_{u}^{n+1},\forall u\}\!\!\bigg),
		\end{align}
	which presents that the value of the objective function exhibits a monotonically non-decreasing trend after each iteration. 
		The inequality marked by $a$ holds because $\mathbf{u}^{n+1}$ represents the optimal receive filter for sensing via step 3 of \textbf{Algorithm \ref{overall algorithm}}.
	    The inequality marked by $b$ holds because $\mathcal{W}_{\mathrm{c}}^{n+1}$ and $\mathbf{V}_{\mathrm{r}}^{n+1}$ represent the optimal receive filter and transmission power design for UL communication via step 4 of \textbf{Algorithm \ref{overall algorithm}}.
		Similarly, the inequality marked by $c$ holds because $\{\mathbf{v}_{u}^{n+1},\forall u\}$ and $\{p_{u}^{n+1},\forall u\}$ represent the optimal transmit beamforming for DL communication and sensing via step 5 of \textbf{Algorithm \ref{overall algorithm}}.
		In addition, the sum rate is upper bounded owing to the finite power available at the PBS and UL UEs. Thus, it confirms that convergence of \textbf{Algorithm \ref{overall algorithm}} is theoretically guaranteed.
        
		\subsubsection{Computational Complexity Analysis}
		It is noted that the computational burden in \textbf{Algorithm \ref{overall algorithm}} mainly 
		results from optimizing $\mathbf{u}$, $\mathcal{W}_{\mathrm{c}}$, $\mathbf{W}_{\mathrm{r}}$, $\{\mathbf{v}_{u},\forall u\}$ and $\{p_{u},\forall u\}$.
	    For updating sensing receive filter $\mathbf{u}$, the complexity lies in the calculation of matrix inversion and  is of order $\mathcal{O}\big((N_0+JN_1)^{3}\big)$.
		For optimizing transmit beamforming $\mathcal{W}_{\mathrm{c}}$ and $\mathbf{V}_{\mathrm{r}}$, the computational complexity is of order $\mathcal{O}\big(I_{\mathrm{o}}I_{\mathrm{i}}\big((DM^2+M^2)^{3.5}+DM^{3.5}\big)\big)$, where $I_{\mathrm{i}}$ and $I_{\mathrm{o}}$ represent the maximum number of inner and outer iterations for convergence, respectively  \cite{S. Boyd}. 
		For optimizing receive filter $\{\mathbf{v}_{u},\forall u\}$ and transmission power $\{p_{u},\forall u\}$ for UL communication, the computational complexity is of order $\mathcal{O}\big(U^{3.5} + U(N_0+JN_1)^{3.5}\big)$. 
		Therefore, the overall computational complexity of \textbf{Algorithm \ref{overall algorithm}} is of order $\mathcal{O}\big(\log(1/\xi)\big((N_0+JN_1)^{3}+I_{\mathrm{o}}I_{\mathrm{i}}\big((DM^2+M^2)^{3.5}+DM^{3.5}\big)+U^{3.5}+ U(N_0+JN_1)^{3.5}\big)\big)$.

		\vspace{-2mm}
		\section{Numerical Results}
		
		In this section, we present numerical results to evaluate the performance of the proposed algorithm for the JAPS-based cooperative multi-static ISAC networks. We conduct 500 Monte Carlo simulations in a 500~m~$\times$~500~m region to validate the generalizability of our algorithm, consisting of one FD ISAC PBS, $J =$ 3 SBSs,  one target, $D =$ 2 DL UEs and $U = $ 2 UL UEs.
        The location of PBS is set to (0 m, 250 m). The PBS is equipped with $M =6$ transmit antennas.
		The PBS and each SBS are equipped with $N_{0}=N_{1} \triangleq N=$~6 receive antennas.
		The target is in the center of the region, while all the UEs and SBSs are deployed randomly relative to the PBS as the reference. 
		The directions of DL UEs $\{\theta_1^{\mathrm{D}},\theta_2^{\mathrm{D}}\}$ and UL UEs $\{\theta_1^{\mathrm{U}},\theta_2^{\mathrm{U}}\}$ are set to $\{-55^\circ,30^\circ\}$ and $\{-70^\circ,20^\circ\}$, respectively.
		The required sensing SINR threshold is set to $\gamma_\mathrm{s} =$ 10 dB. 
		The maximum transmit power at the PBS is set to  $P_{\mathrm{max}}^{\mathrm{PBS}} =$ 30~dBm,  while the maximum transmission power at each UL UE is set to $P_u^{\mathrm{max}} =$ 16~dBm,$\forall u$.
		The Rician factors are set as 3~dB.  Based on the typical distance-dependent path loss model \cite{liurang SNR,wuqingqing RIS2}, we set $C_0=-$30~dB. Moreover, we assume the channel path-loss exponents for PBS-target, target-SBSs,  PBS-UEs and UE-SBSs links are set as 2.3, 2.3, 2.4 and 2.5, respectively. For simplicity, we assume the residual SI gain is $\beta_{\mathrm{SI}} =-$110~dB \cite{youxiaohu DL UL,liurang SNR}. 
        Without losing generality, we assume the combined sensing channel gains are $\alpha_\iota=\frac{\sigma_\iota}{2d_\iota}$ \cite{liufan RCS,AISAC},  where $d_\iota$ denotes the distance between the PBS/SBS $j$ and target. $\sigma_\iota$ represents the complex RCS,  which follows Swerling-I model, and the probability density function of RCS $\sigma$ satisfies
$f(\sigma_\iota)=\frac{1}{\sigma_0}\exp(-\frac{\sigma_\iota}{\sigma_0}),\sigma_\iota\geq0$, where $\sigma_0$ is the average value of target’s RCS \cite{luohongliang Clutter Environment}. The noise powers for communication and sensing are set as $-$80~dBm.
		In addition, the initial penalty factor can be set to $\eta_1=10^{4}$. 
       The scaling coefficient $\epsilon_1$ is set as $\epsilon_1=0.7$.
        The convergence thresholds of inner and outer layers for optimizing transmit beamforming can be set to  $\varepsilon_1=10^{-2}$ and $\varepsilon_2=10^{-5}$, respectively.
		Finally,  we set the convergence threshold of our proposed algorithm as $\xi=~10^{-4}$. 
\begin{figure}[t]
	\centering
	\includegraphics[width=68mm]{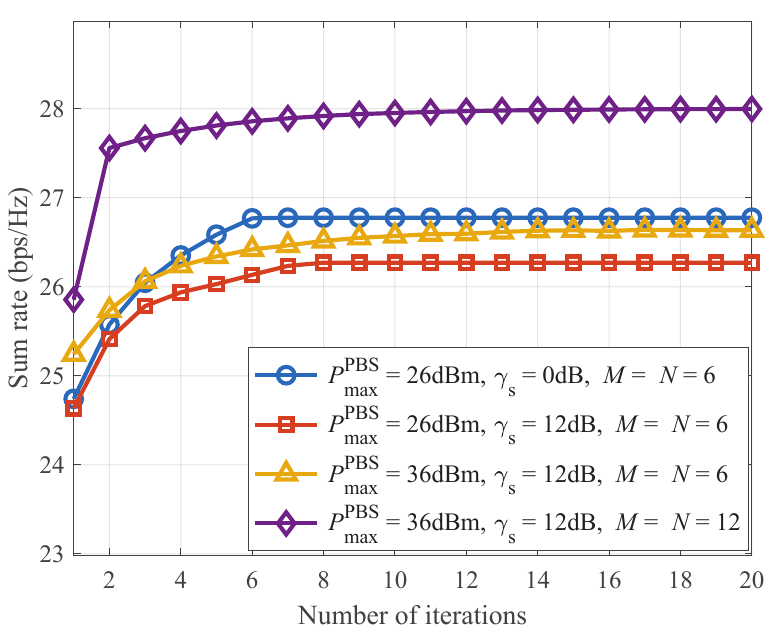}
	\caption{Convergence behavior of the proposed algorithm.}
	\label{fig:Convergence performance}
\end{figure}
		
		We first illustrate the convergence performance of our proposed optimization algorithm in Fig.~\ref{fig:Convergence performance}, which shows the variation of multi-UE sum rate as the number of iterations under different PBS transmit power $P_{\mathrm{max}}^{\mathrm{PBS}}$, different sensing requirements $\gamma_\mathrm{s}$  and different antenna numbers.
		It can be observed that the multi-UE sum rate grows rapidly with higher iteration numbers and can converge within about 10 iterations under different settings, which indicates the effectiveness and advantages of our proposed algorithm. 
	
\begin{figure}[t]
		\centering
		\includegraphics[width=68mm]{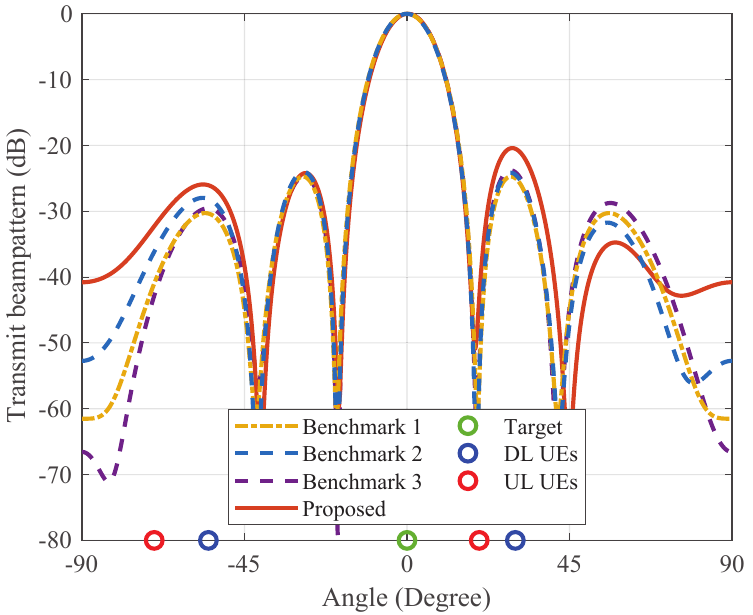}
		\caption{PBS transmit beampattern of the proposed algorithm.}
		\label{fig:transmit_beampartten}
	\end{figure}

	Next, we demonstrate the PBS transmit beampattern gain which is attained through our proposed algorithm compared with the other benchmark algorithms as follows. (1) Benchmark 1 (i.e., the regularized zero-forcing (RZF) beamforming) \cite{Integrated Active and Passive Sensing}: This scheme employs the RZF transmit beamforming instead of the optimal transmit beamforming in Section \ref{Transmit Beamforming Design} used in the proposed algorithm.
    Zero-forcing (ZF) methods are relatively simple, facilitate the acquisition of closed-form and interpretable precoders, and have the potential for near-optimal performance in multi-UE communication systems \cite{liuxiang ISAC}. (2) Benchmark 2 (i.e., the detection probability maximization)  \cite{IAPS}: This scheme maximizes the detection probability of the target while satisfying the constraint \eqref{eq:power constraint P0} and the minimum SINR demands for DL and UL communications.
    (3) Benchmark 3 (i.e., sensing SINR maximization): In this case, we wish to maximize sensing SINR with the same  transmit power budget at the PBS, as well as DL and UL communications constraints. 
	Fig.~\ref{fig:transmit_beampartten} shows that the PBS transmit beams are pointed towards the target and two DL UEs, respectively.
	Obviously, we observe that the transmit beampattern gain of the proposed algorithm surpasses that of other benchmark schemes.
	This is because the goal of our proposed algorithm is to maximize the multi-UE sum rate, which allows the available transmission power to be fully used for communication.
		
	\begin{figure}[t]
		\centering
		\includegraphics[width=68mm]{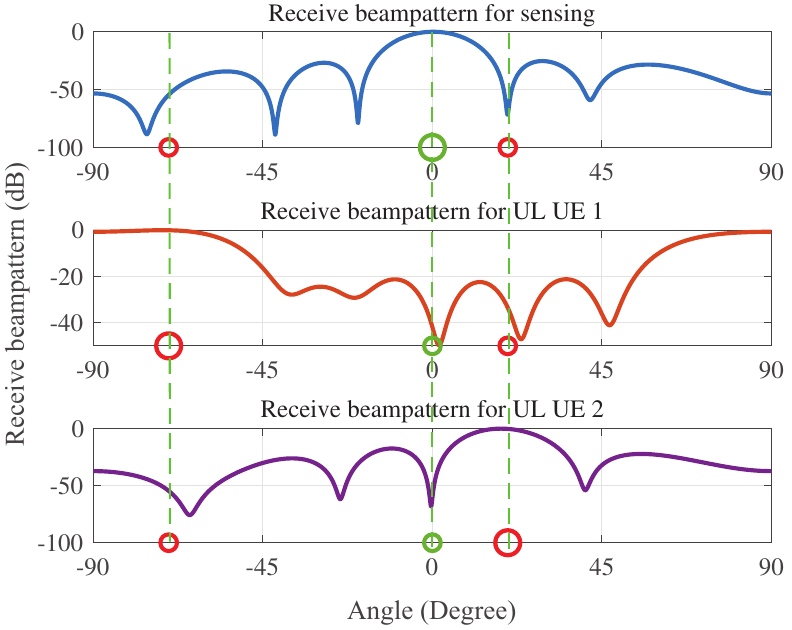}
		\caption{PBS receive beampattern of the proposed algorithm (The green circle indicates the direction of the target, and the red circles indicate the directions of UL UEs. The larger circle represents the expected direction of the receive beam).}
		\label{fig:receive_beampartten}
	\end{figure}

		Then, we show the PBS receive beampattern gains for sensing and the UL communication functionality respectively, as depicted in Fig.~\ref{fig:receive_beampartten}. From the first subfigure, a main beam is allocated to point at the target for target detection.
		Meanwhile, since the signals transmitted by two UL UEs cause substantial interference to target sensing, several deep nulls are steered toward them.
		Similar observations can be seen in the second subfigure and the third subfigure in Fig.~\ref{fig:receive_beampartten}. For one UL UE, the corresponding main receive beam is pointed towards it, while the corresponding relatively deep nulls are steered toward the directions of target and the other UL UE. 
		It is worth recalling that two main beams of the transmitted signal are oriented towards the corresponding DL UEs in Fig.~\ref{fig:transmit_beampartten}.
		Combining these two facts, we can conclude that the proposed algorithm excels in ensuring reliable communication functionality.
		
		\begin{figure}[t]
			\centering
			\includegraphics[width=68mm]{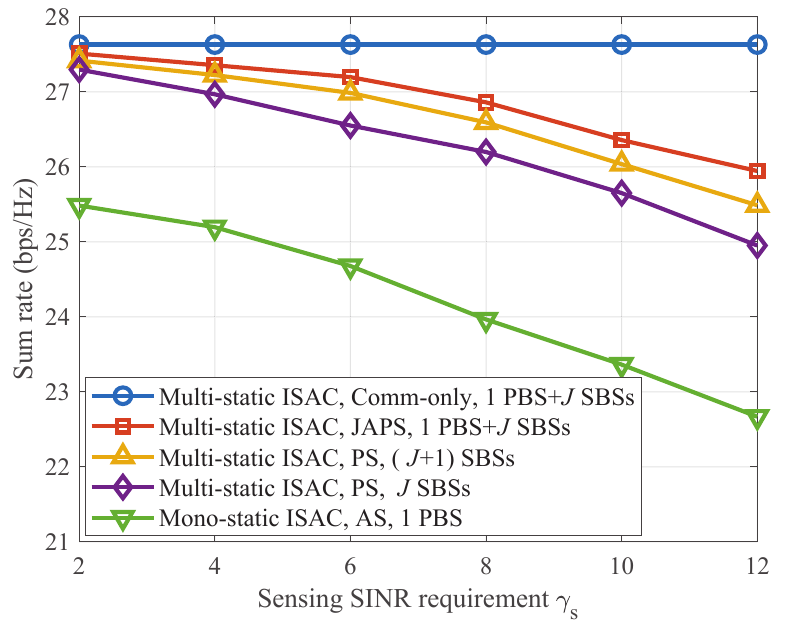}
			\caption{Multi-UE sum rate versus the sensing SINR threshold $\gamma_\mathrm{s}$.}
			\label{fig:R_SINR}
		\end{figure}
		
		\begin{figure}[t]
			\centering
			\includegraphics[width=68mm]{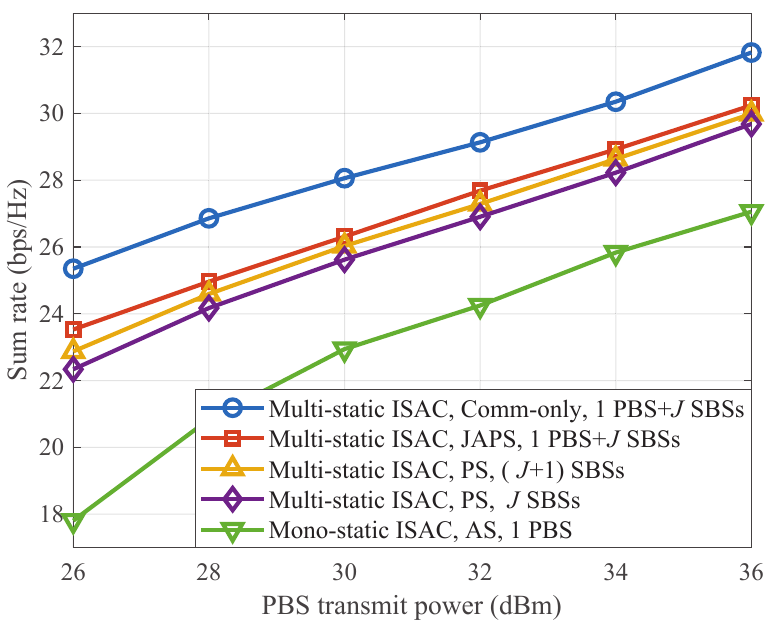}
			\caption{Multi-UE sum rate versus the PBS transmit power budget $P_{\mathrm{max}}^{\mathrm{PBS}}$.}
			\label{fig:R_Pmax}
		\end{figure}
		
		To better verify the advantage of the proposed algorithm in the JAPS-based cooperative multi-static ISAC network (i.e., JAPS), active-only and passive-only sensing schemes (i.e., AS and PS, respectively) are provided for comparison.
        It is worth emphasizing that both the proposed JAPS scheme and the PS scheme are cooperative multi-static ISAC modes. The proposed JAPS scheme uses the PBS and other SBSs to simultaneously receive echo signals and UL communication signals, while the PS scheme uses all SBSs to receive echo signals and UL communication signals. 
        In contrast, the AS scheme is a mono-static ISAC mode, where only the PBS receives echo signals and UL communication signals. 
        To ensure fairness in comparison, we further evaluate the performance of the PS scheme equipped with $J$ SBSs and the PS scheme equipped with $(J+1)$ SBSs, respectively. Besides, the communication-only scheme (i.e., Comm-only) without considering the sensing constraint is also included as the benchmark to present the upper bound of the multi-UE sum rate performance for the considered ISAC system.

		The multi-UE sum rate versus sensing SINR requirement $\gamma_\mathrm{s}$ is shown in Fig.~\ref{fig:R_SINR}. 
		A higher sensing SINR requirement $\gamma_\mathrm{s}$  diminishes the multi-UE sum rate in the ISAC system, illustrating the trade-off between sensing and multi-UE communications. 
		Besides, the proposed JAPS scheme is less sensitive to the value of sensing SINR requirement $\gamma_\mathrm{s}$ compared to the benchmark AS and PS schemes, which means the proposed JAPS scheme has better sensing tolerance.

		In Fig.~\ref{fig:R_Pmax}, we investigate the variation of multi-UE sum rate with maximum PBS transmit power $P_{\mathrm{max}}^{\mathrm{PBS}}$. Augmenting the maximum PBS transmit power yields progressive improvements in the system sum rate.
		Larger PBS transmit power $P_{\mathrm{max}}^{\mathrm{PBS}}$ provides larger beamforming gains since more resources can be exploited. 
		It is unequivocal that the Comm-only scheme demonstrates the most superior multi-UE communication sum rate performance.
		Moreover, the results demonstrate that the proposed JAPS scheme exhibits notable performance gains over both AS and PS baselines,  evidencing the superiority of our proposed JAPS scheme.
		\begin{figure}[t!]
			\centering  
			\subfigure[Circular topology]{   
				\begin{minipage}{4.1cm}
					\centering    
					\includegraphics[height=40mm]{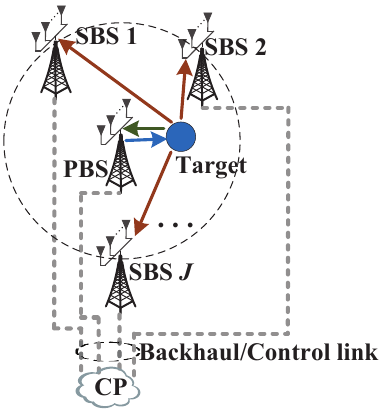}  
				\end{minipage}
			}
			\subfigure[Linear topology]{ 
				\begin{minipage}{4.1cm}
					\centering    
					\includegraphics[height=40mm]{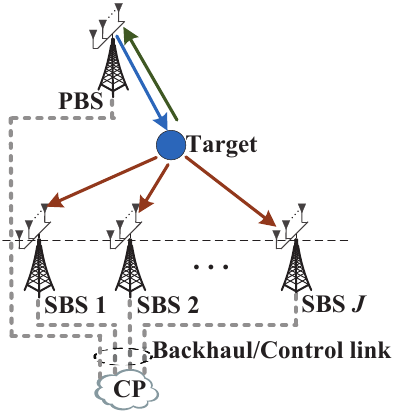}
				\end{minipage}
			}
			\caption{A case with circular and linear topology of the SBSs.}   
			\label{fig:circular and linear}    
		\end{figure}
		
		\begin{figure}[t!]
			\centering
			\includegraphics[width=68mm]{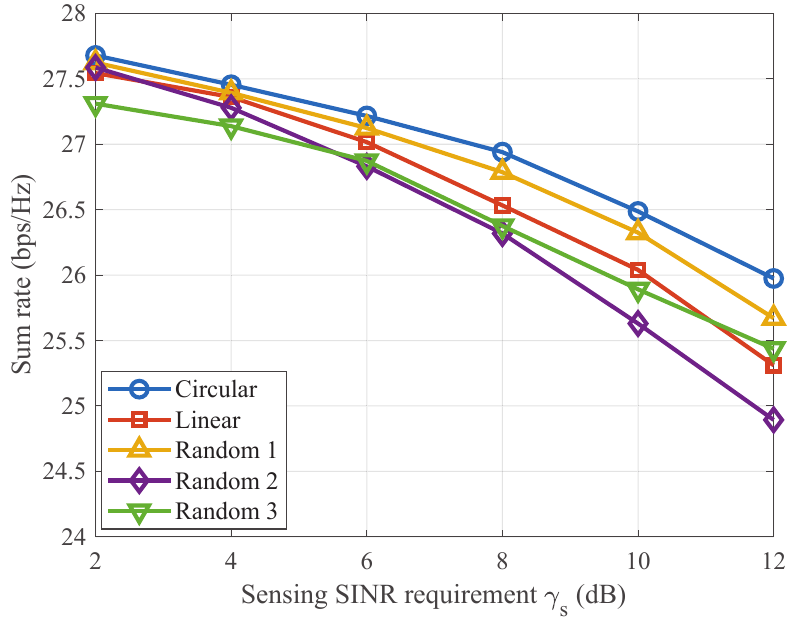}
			\caption{Performance comparison under different topologies.}
			\label{fig:different_topologies_SNR}
		\end{figure}

To further explore the effect of distribution of SBSs on system performance, we illustrate in Fig.~\ref{fig:different_topologies_SNR} the multi-UE sum rate of the proposed JAPS scheme versus sensing SINR requirement $\gamma_\mathrm{s}$ under different SBS topologies, which contain three topologies: the circular topology (applicable for security scenarios in important facilities, such as airports and nuclear power plants), the linear topology (applicable for linear area surveillance, such as borders, coastlines, railways and highway corridors), and the random topology \cite{liuan BS fenbu}. Fig.~\ref{fig:circular and linear} is a diagram of the circular and linear topology of the SBSs.		
In the circular topology, SBSs are symmetrically distributed along a 200 m radius circle centered on the PBS, with the target uniformly distributed within the circle. For the linear topology, SBSs are linearly distributed along a straight line with 60 m spacing, while the PBS is positioned 200 m from this line. The target is uniformly distributed along a parallel line situated 80 m from the SBS line.
It is observed that the system performance under the circle topology is superior to that under the linear topology and other random topologies, which suggests that the SBSs topology should be carefully considered in practice.
		
		\begin{figure}[t]
			\centering
			\includegraphics[width=68mm]{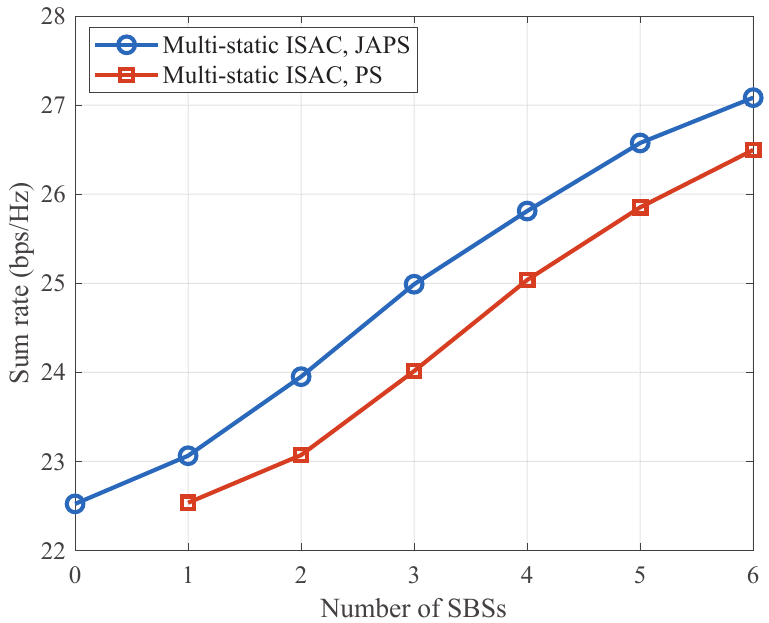}
			\caption{Multi-UE sum rate versus the number of SBSs.}
			\label{fig:R_SBS}
		\end{figure}

		Fig.~\ref{fig:R_SBS} shows the multi-UE sum rate versus the number of SBSs. We find that the multi-UE sum rate for both our proposed JAPS scheme and PS scheme increases gradually as the deployment of SBSs grows.
	This is because the CP gains more sensing information with a growing number of SBSs, which facilitates the system in satisfying the sensing SINR constraint. Furthermore, more resources can be available to enhance communication quality. 
		In addition, our proposed JAPS scheme demonstrates superior performance compared to PS scheme.
\begin{figure}[t]
			\centering
			\includegraphics[width=68mm]{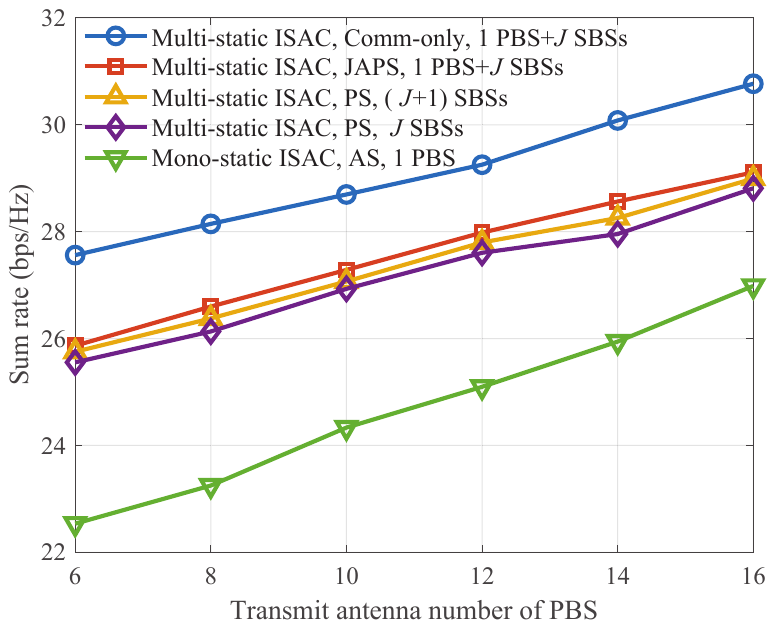}
			\caption{Multi-UE sum rate versus different PBS transmit antenna numbers.}
			\label{fig:R_PBS_M}
		\end{figure}

\begin{figure}[t]
			\centering
			\includegraphics[width=68mm]{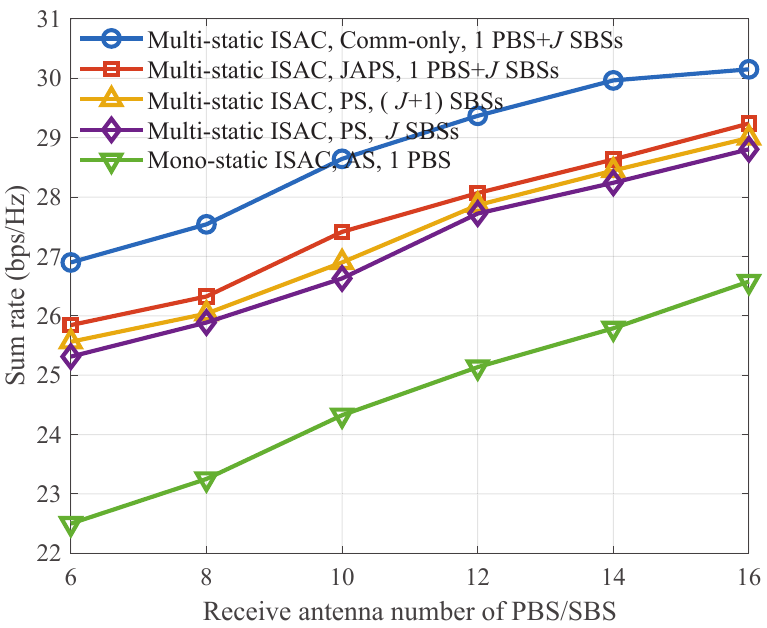}
			\caption{Multi-UE sum rate versus different PBS/SBSs receive antenna numbers.}
			\label{fig:R_PBS_SBS_N}
\end{figure}
		
		Finally, Fig.~\ref{fig:R_PBS_M} and Fig.~\ref{fig:R_PBS_SBS_N} illustrate the relationship between multi-UE sum rate and the antenna numbers under different schemes.
		The multi-UE sum rate achieved by our proposed JAPS scheme is higher than that achieved by PS and AS schemes.
		it is clear that the increase of the number of transmit and receive antennas has an obvious effect on improving system performance. This is because more antennas can expand spatial DoFs while enabling higher combining gains through beamforming optimization.
		Together with the fact in Fig.~\ref{fig:R_Pmax}, the performance of the ISAC networks has been significantly improved with the increase of resources.
		
		\section{Conclusion}
		
		In this paper, a unified design framework for active and passive sensing has been proposed.
		We have investigated the joint beamforming design and power optimization for the JAPS-based cooperative multi-static ISAC system for coexisting UL and DL communications.  Specifically, the sum rate for multi-UE communications has been maximized while adhering to the sensing SINR requirement, transmit power budget at the PBS and UL UEs. 
		First, to deal with the resulting complicated problem, the primal problem has been decoupled into three sub-problems. Given the other variables, we have applied SCA-based,  penalty-based and FP-based iterative algorithms to optimize these subproblems alternately until convergence.
		Next, the convergence of the proposed algorithm has been analyzed, and its computational complexity has been derived.
	    Finally, numerical results have been provided to validate the convergence and effectiveness of our proposed algorithm, and illustrated the performance improvements introduced by our proposed unified design framework for JAPS. Furthermore, numerical results have also demonstrated the obvious superiority of our proposed algorithm in interference mitigation.
        In our future research, we will further investigate more meaningful extensions in cooperative multi-static ISAC networks, such as robust design methodologies under imperfect channel state information and secure transmission strategies aimed at preventing unintended information leakage.

	\end{document}